\documentclass[useAMS,usenatbib]{mn2e}
\usepackage{rotating}
\usepackage{subfigure}
\RequirePackage{setspace}
\usepackage{graphicx}
\usepackage{fancyhdr}
\usepackage{natbib}
\usepackage{amsmath}
\usepackage{amssymb}
\usepackage{aas_macros}
\usepackage{epstopdf}
\usepackage{lscape}
\usepackage{rotating}
\usepackage[T1]{fontenc}
\font\btt=rm-lmtk10

\title[The G10/COSMOS region]{Galaxy And Mass Assembly (GAMA): Curation and reanalysis of 16.6k redshifts in the G10/COSMOS region}
\author[L. J. M. Davies et. al.]{L. J. M. Davies$^{1}$ \thanks{E-mail:
luke.j.davies@uwa.edu.au}, S. P. Driver$^{1,2}$, A. S. G. Robotham$^{1}$, I. K. Baldry$^{3}$, \newauthor R. Lange$^{1}$, J. Liske$^{4}$,  M. Meyer$^{1}$, A. Popping$^{1}$, S. M. Wilkins$^{5}$, A. H. Wright$^{1}$\\
$^{1}$ICRAR, The University of Western Australia, 35 Stirling Highway, Crawley, WA 6009, Australia \\
$^{2}$SUPA, School of Physics \& Astronomy, University of St Andrews, North Haugh, St Andrews, KY16 9SS, U.K. \\
$^{3}$Astrophysics Research Institute, Liverpool John Moores University, IC2, Liverpool Science Park, 146 Brownlow Hill, Liverpool L3 5RF, U.K. \\
$^{4}$European Southern Observatory, Karl-Schwarzschild-Str. 2, 85748 Garching, Germany\\
$^{5}$Astronomy Centre, Department of Physics and Astronomy, University of Sussex, Brighton, BN1 9QH, U.K. \\
}

\begin{document}

\date{Draft: Sept 2014}

\pagerange{\pageref{firstpage}--\pageref{lastpage}} \pubyear{2014}

\maketitle

\begin{abstract}

We discuss the construction of the Galaxy And Mass Assembly (GAMA) 10$^\mathrm{h}$ region (G10) using publicly available data in the Cosmic Evolution Survey region (COSMOS) in order to extend the GAMA survey to $z\sim1$ in a single $\sim$1\,deg$^{2}$ field. In order to obtain the maximum number of high precision spectroscopic redshifts we re-reduce all archival zCOSMOS-bright data and use the GAMA automatic cross-correlation redshift fitting code \textsc{autoz}. 

We use all available redshift information (\textsc{autoz}, zCOSMOS-bright 10k, PRIMUS, VVDS, SDSS and photometric redshifts) to calculate robust best-fit redshifts for all galaxies and visually inspect all 1D and 2D spectra to obtain 16,583 robust redshifts in the full COSMOS region. We then define the G10 region to be the central  $\sim$1\,deg$^{2}$ of COSMOS, which has relatively high spectroscopic completeness, and encompasses the CHILES VLA region. We define a combined $r<23.0$\,mag \& $i<22.0$\,mag G10 sample (selected to have the highest bijective overlap) with which to perform future analysis, containing 9,861 sources with reliable high precision VLT-VIMOS spectra. All tables, spectra and imaging are available at: {\btt http://ict.icrar.org/cutout/G10}.

\end{abstract}


\section{Introduction}
\label{sec:intro}

The GAMA survey is a highly complete multi-wavelength database \citep{Driver11, Liske14} and galaxy redshift ($z$) survey \citep{Baldry10,Robotham10,Hopkins13} covering 280\,deg$^{2}$ to a main survey limit of $r<19.8$\,mag in three equatorial (G09, G12 and G15) and two southern (G02 and G23) regions. The spectroscopic survey was undertaken using the AAOmega fibre-fed spectrograph in conjunction with the Two-degree Field (2dF) positioner on the Anglo-Australian Telescope and contains $\sim$217,000 reliable redshifts covering $0<z\lesssim0.5$ with a median redshift of $z\sim0.2$. The survey, and associated products, have provided a wealth of information regarding galaxy properties \citep[$e.g.$][]{Baldry12,Robotham13,Taylor11}, large scale structure \citep[$e.g.$][]{Alpaslan14}, the cosmic spectral energy distribution \citep[$e.g.$][]{Driver12,Kelvin14}, the role of environment in galaxy evolution \citep[$e.g.$][]{Brough13,Robotham13}, and the spatial distribution and properties of intermediate mass halos \citep[$e.g.$][]{Robotham11,Robotham12,Alpaslan12} to mention but a few key science projects. However, by its construction, the survey is necessarily limited to $z<0.5$ systems. Little is known about the early time evolution of GAMA derived properties as we are essentially only, albeit comprehensively, probing these quantities over the last $\sim4$\,Gyrs. As such, it would greatly benefit the survey to obtain a high redshift benchmark with which to compare to the extensive low redshift data set.

While there are a number of spectroscopic surveys, and associated analysis, which extend to high redshift ($e.g.$ VVDS, VIPERS, DEEP2, zCOSMOS, PRIMUS - see below for details), in their current state, the variety of input catalogue selections, redshift identifications and subsequent analysis renders a direct comparison with GAMA problematic. However, it is possible to use these existing data sets to produce a GAMA comparable high redshift sample. 

The Cosmic Evolution Survey region \citep[COSMOS][]{Scoville07} is ideally suited to this task, with extensive multi-wavelength broadband coverage from X-ray to radio wavelengths, including deep Hubble Space Telescope imaging over the full 1.8\,deg$^{2}$ region \citep{Scoville07}, and deep NIR data from the UltraVISTA survey \citep{McCracken12}, numerous spectroscopic campaigns and a comprehensive photometric redshift catalogue \citep{Ilbert09}. In combination these data can be used to produce a highly complete, magnitude limited, sample of sources in the region with robust spectroscopic redshifts. Essentially we can perform a comprehensive analysis of all available redshift information in the region to produce an accurate redshift catalogue, using the most robust and highest precision redshift available for each source. Clearly, high resolution spectra are desirable and are in fact essential for analyses such as the identification of intermediate mass halos \citep[$e.g.$][]{Robotham11}. However, lower resolution redshifts, if robust, are likely to be adequate for a significant amount of the GAMA-type analysis, such as structural decomposition and mass-size evolution.  

In addition to the extensive multi-wavelength data already available, deep Karl G. Jansky Very Large Array (JVLA) 21-cm observations are currently underway in the COSMOS regions as part of the COSMOS HI Large Extragalactic Survey \cite[CHILES see,][]{Fernandez13}. CHILES is an SKA pathfinder survey which will target 21cm HI emission at $0<z\lesssim0.45$ in a single $\sim$0.3\,deg$^{2}$ FWHM pointing in the COSMOS region. Forming a robust, magnitude limited spectroscopic sample in the CHILES region will allow the CHILES data to be fully exploited, leading to deep HI stacking of optically selected sources in order to constrain the low gas mass end of the HI mass function. In addition, comprehensive optically selected samples provide an accurate quantification of local environment through large-scale-structure, group catalogues and the local galaxy density. HI is a poor tracer of environment since it is anti-biased due to gas stripping in the highest density regions. As such, optical samples are required to constrain the environment of HI detected galaxies helping to explain the seemingly contradictory results of the environmental dependance on the HI mass function \citep[$e.g.$][]{Springob05, Zwaan05}.     

 In conjunction with SED derived stellar masses (such as those produced by GAMA and ultimately the sample here), the CHILES data and spectroscopic catalogue defined in this work will allow a complete analysis of the baryonic mass function and stellar/gas mass ratios in a significantly large sample of galaxies.    

While the majority of the COSMOS region data has been made publicly available in a reduced, immediately useable format, one notable exception is the ESO Large-program spectroscopic survey, zCOSMOS \citep{Lilly07} - undertaken using the VISible Multi-Object Spectrograph (VIMOS). The zCOSMOS survey targeted a total of $\sim30,000$ sources in the COSMOS region split into a `bright' survey (targeting 20,000 I-band selected sources at $z<1.2$) and a `deep' survey (targeting 10,000 colour-selected sources at $1.5<z<3$). To date, the zCOSMOS team have only publicly released 10,109 redshifts from the `bright' survey \citep[the 10k release,][]{Lilly09}. In this paper we re-reduce the entire zCOSMOS-bright data and provide the redshifts which were not publicly released as part of the the zCOSMOS-bright 10k sample  - full details of the zCOSMOS survey can be found in \citet{Lilly07,Lilly09} and are summarised in Section \ref{sec:spec_reduc}. 

In addition to the zCOSMOS data set, there are a number of other spectroscopic surveys in the COSMOS region. These include the VIMOS VLT Deep Survey \ \citep[VVDS,][]{Garilli08}, Sloan Digital Sky Survey DR10 \citep[SDSS,][]{Ahn14} and the PRIsm MUlti-object Survey \citep[PRIMUS\footnote{For PRIMUS public data release see {\btt http://primus.ucsd.edu}},][]{Coil11,Cool13}, a spectroscopic survey of $\sim$120,000 galaxies covering 9\,deg$^{2}$ out to $z\sim1$, $\sim$30,000 of which are in the COSMOS region. The spectral resolution of PRIMUS spectra is somewhat lower than zCOSMOS ($\sim$1500\,km\,s$^{-1}$ or R$\sim$20), but as noted above, is adequate for the bulk of our future analysis. 

In this paper we construct an $r$-band and $i$-band magnitude limited sample in a sub region of COSMOS (the G10 region) compiled from the zCOSMOS-bright 10k sample, PRIMUS, VVDS, SDSS and our own zCOSMOS re-reduction and analysis. Where spectroscopic redshifts are not available we supplement our final catalogue with the accurate (30-band) photometric redshifts of \cite{Ilbert09}.  This paper is structured as follows: In Section \ref{sec:spec_reduc} we discuss the re-reduction of all archival zCOSMOS-bright data and outline the automatic redshift fitting using \textsc{autoz}, in Section \ref{sec:match} we detail the position matching of spectroscopic targets to photometric sources, in Section \ref{sec:best_red} we detail the selection of the most robust redshift for each source using all available redshift data, and in Section \ref{sec:sample} we discuss the properties of the final G10 spectroscopic sample and its uses as a high redshift extension of the GAMA survey. Throughout this paper all magnitudes are on the AB scale, and the cosmology used is \textit{H}$_{0}$=100\,kms$^{-1}$ Mpc$^{-1}$, $h=H_0/100$\,kms$^{-1}$ Mpc$^{-1}$, $\Omega_{\Lambda}$ = 0.7 and $\Omega_{M}$ =0.3.

All data from this analysis, as well as a tool for displaying photometric data associated with each source in the COSMOS region is made publicly available (see Appendix for details).

\begin{figure}
\begin{center}
\includegraphics[scale=0.4]{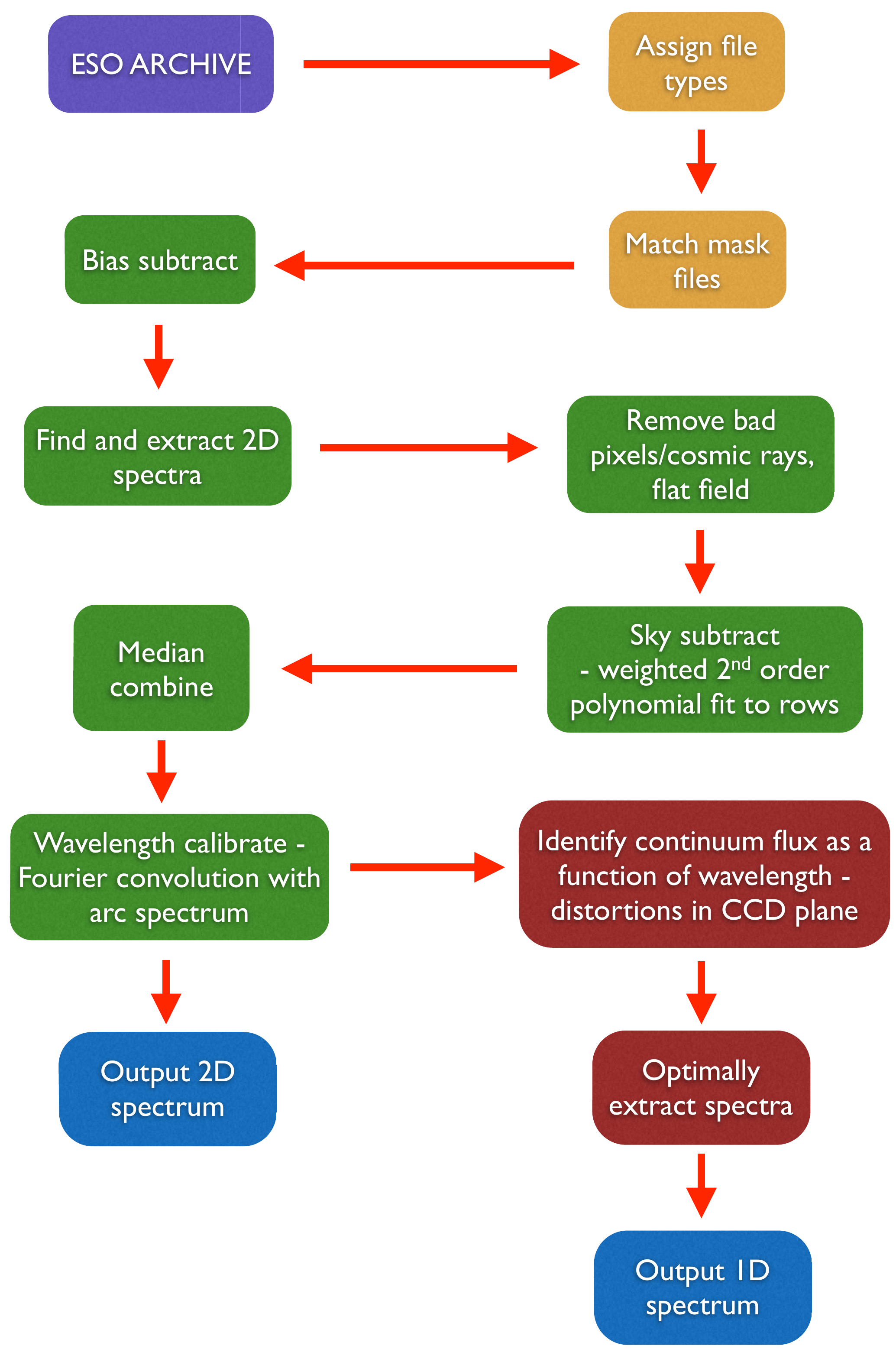}

\caption{Flow chart detailing the key processes of the R pipeline used to re-reduce all zCOSMOS data. Colours represent different stages in the reduction process: data acquisition (purple), pre-pipeline organisation (orange), data reduction (green), post-reduction spectral extraction (brown) and pipeline out-puts (blue). }
\label{fig:flow}
\end{center}
\end{figure}

\section{Re-reduction of zCOSMOS data and  redshifting }
\label{sec:spec_reduc}

\subsection{The bespoke R reduction pipeline}

To re-reduce all archival zCOSMOS-bright data we produced a bespoke R pipeline for the reduction of VIMOS spectroscopy. This pipeline is similar to that used in the redshift survey towards the CMB cold spot \citep{Bremer10}. Briefly, all zCOSMOS-bright data and calibration files were downloaded from the ESO archive (PID:175.A-0839, P.I. Lily), then reduced and calibrated in a fully automated process (see Figure \ref{fig:flow}). 

Initially all science frames were matched with the correct calibration files using their mask number and observation date. We then identified all spectra in each individual science and flat field frame by taking a cross section through the image and identifying regions of the frame which are above the background level (the fits header slit positions are found to be inaccurate and vary in offset with the true positions as a function of observation time and frame position). We extracted each 2D spectrum, removed bad pixels and cosmic rays, and flat fielded all raw science frames. Next we performed a sky subtraction by fitting a weighted second order polynomial fit to each spectrum row, perpendicular to the dispersion axis (weightings are used to exclude regions of source flux in the background estimation). zCOSMOS spectra were obtained using a `jitter along the slit' technique, where the target source is positioned at different pixel positions along the slit during each exposure. We corrected for jitter offsets and median combined all spectra. Each combined spectrum is then wavelength calibrated by convolving the inverse Fourier transforms of an arc lamp spectrum and known arc lamp emission line list. The VIMOS instrument was known to suffer from flexure issues during the zCOSMOS observations (which have now been improved), these problems can cause variable wavelength offsets along the dispersion axis between individual exposures. This distortions are non linear and as such can not be corrected for by a rigid offset. In previous iterations of our reduction pipeline we have taken the time consuming approach of wavelength calibrating individual exposures, and correcting each spectrum for wavelength distortions prior to combination. However, in the zCOSMOS observations, we found the differences between wavelength distortions over individual exposures to be small (a 2 pixel offset at most), and to be most pronounced at the very red end of each spectrum - where emission lines are strongly contaminated by skylines and are largely unusable. As such, we do not correct for this effect, but do allow for non-linear wavelength solutions in our final wavelength calibration. We note that for a small number of cases this marginally reduces our redshift accuracy, but will not significantly alter our results.

At this stage the pipeline outputs a reduced two-dimensional spectrum. We then median compressed each spectrum along the dispersion axis and identified the source continuum flux position along the spatial axis. We traced the continuum position of each source along the dispersion axis in order to identify any curvature in the spectrum on the CCD and performed an optimal extraction of source flux; tracing the source curvature and identifying pixels in each row which have flux  $>1.5\sigma$ of the background level.  Lastly, we output wavelength calibrated, optimally extracted one-dimensional spectra for each zCOSMOS-bright target source. 

\begin{figure*}
\begin{center}
\includegraphics[scale=0.37]{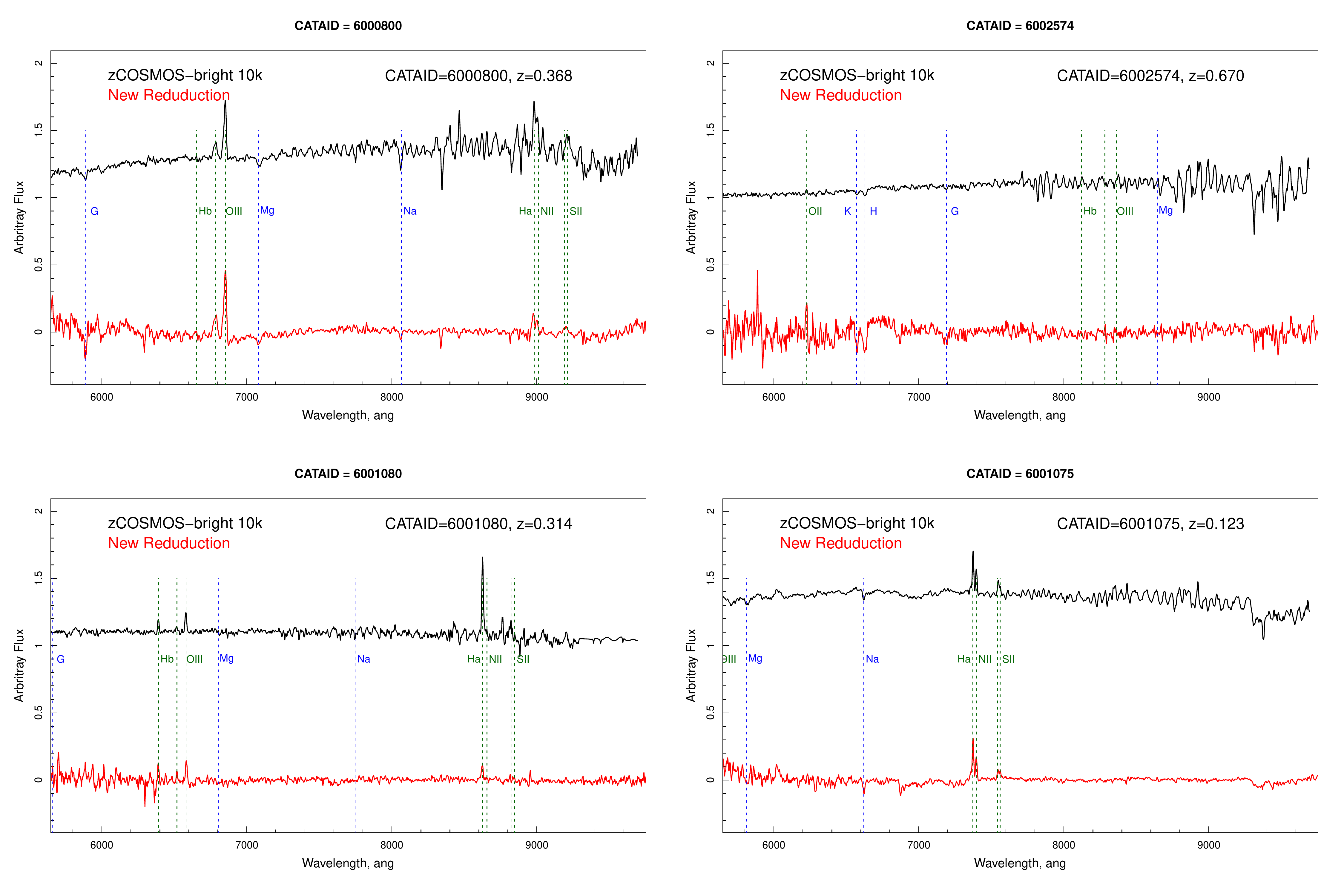}
\caption{Comparison of our data reduction method (red line) with the original 1D spectrum from the zCOSMOS-bright 10k release (black line). We over-plot key spectral emission (green dashed vertical lines) and absorption (blue dashed vertical lines) features used for obtaining galaxy redshifts.}
\label{fig:specs}
\end{center}
\end{figure*}

The \textsc{autoz} fitting code \citep[][] {Baldry14} does not use spectral shape in its redshift fitting, only emission and absorption features. Hence, our pipeline is designed to maximise signal to noise in spectral features and does not conserve spectral shape or flux. Four randomly selected examples of the final one-dimensional spectra from our reduction, can be seen in Figure \ref{fig:specs}. We find that our re-reduction of the zCOSMOS-bright data produces spectra which are largely consistent in signal to noise to the zCOSMOS-bright 10k release spectra. In some cases we find slightly higher signal emission line features (as in GAMA catalogue ID, CATAID=6000800), while in other cases we find the signal to noise of emission line features slightly reduced (as in CATAID=6001080). However, in using the original zCOSMOS-bright publicly released data in combination with our new reduction, we can hope to improve the number of galaxies with reliable redshifts in the COSMOS region (if only those which benefit from our new reduction method).

We found that for $\sim$1000 of the zCOSMOS-bright spectra our pipeline failed to successfully extract a one-dimensional spectrum. This was generally due to one of four problems: i) the target source spectrum was heavily distorted on the CCD - causing the continuum to fall at the large angle across the spectrum and not entirely lie within the extracted 2D spectrum, ii) the target source continuum was extremely close to the edge of the extracted 2D spectrum and fell partially off the edge, iii) fringing in the sky lines at the red end of the spectrum were extremely strong and source continuum could not be identified and iv) no source continuum could be identified but there were possible faint emission line features. In these cases we visually inspected the 2D spectra for each source and manually extracted a 1D spectrum, allowing for curvature.

\subsection{Automatic redshift fitting using \textsc{autoz}}

In order to obtain redshifts from our independent zCOSMOS-bright data reduction we used an adapted version of the automatic redshift fitting code developed for GAMA spectroscopy, \textsc{autoz} \citep{Baldry14}. Briefly, \textsc{autoz} continuum subtracts the target spectra and cross correlates with a sample of continuum subtracted galaxy templates to identify both absorption-line and emission-line features. In this process, deviations in the high-pass filtered spectra are clipped to exclude uncorrected artefacts and to reduce the significance given to single-line matches. For full details of the \textsc{autoz} code see \cite{Baldry14}. For use with our zCOSMOS data we applied a wrapper to the \textsc{autoz} code to perform additional skyline masking and to smooth the galaxy templates to zCOSMOS scales. If our initial best-fit \textsc{autoz} redshift is consistent with mis-identified sky lines (as happens frequently for single line redshifts at $z>0.7$ in the zCOSMOS data), we use more heavily masked skylines in our reduced spectra and repeated our \textsc{autoz} analysis. \textsc{autoz} was run over all $\sim$20,000 sources from our zCOSMOS reduction, details of how these redshifts were used in the construction of the final G10 sample are given in the following section. Note that we do not use the redshift confidences provided by \textsc{autoz} in this work as they are calibrated for a specific telescope and instrumental setup. However, we do provide confidences for all \textsc{autoz} fits in our final G10 catalogue.

\section{Position matching and catalogue construction}
\label{sec:match}

To build our catalogue we initially matched the 2007 photometric catalogue described in \cite{Capak07} to the updated 2008 COSMOS catalogue including the deep Subaru observations of \cite{Taniguchi07}, to obtain the optical-NIR photometric data available for each source. We then matched each photometric source to the accurate 30-band photometric redshift catalogues of \cite{Ilbert09}. These catalogues all contain COSMOS survey IDs and as such, do not require position matching. Following this, we matched the zCOSMOS-10k bright \citep[][]{Lilly09}, PRIMUS \citep[][]{Coil11,Cool13}, VVDS \citep[][]{Garilli08} and SDSS-DR10 \citep[][]{Ahn14} sources to the photometrically identified sources using a 2$^{\prime\prime}$ match criteria - retaining quality flags for all sources.      

While the zCOSMOS-bright raw data is publicly available through the ESO archive, we do not have access to the input catalogues of targets selected for the VIMOS observations. The zCOSMOS raw data contains the Multi-Object Spectrograph (MOS) slit position, but there is ambiguity between the slit position and the true sources targeted by the observation. In addition, multiple sources may fall in the same slit and hence, we must discern which photometric object relates to our re-reduced spectrum.  We reconstructed the zCOSMOS-bright input catalogue using the details outlined in \citep{Lilly07}. We selected sources using the HST F814 magnitude, where available, and Canada France Hawaii Telescope (CFHT) $i$-band observation for regions without HST data, with a liberal criterion of $i<23.5$\,mag (the zCOSMOS input catalogue is defined at $i<22.5$\,mag  but is stated to contain a number of sources at $i$>$22.5$\,mag ). Both magnitudes were obtained from the \cite{Capak07} catalogue used in the zCOSMOS input selection.

During the compilation of this catalogue we applied multiple position matching algorithms to match the raw data spectra positions to the photometric catalogue. However, due to the ambiguity between slit and true source positions, all algorithms resulted in a low efficiency of true source matching when visually inspected. In order to over come this, we visually inspected $all$ position matches and assigned the best visual match. In this process we initially estimated the source position within each slit by identifying the continuum emission offset from the slit centre and matched to the nearest $i<22.5$\,mag source. We then visually identified the best matched $i<22.5$\,mag source to each slit - taking into account all surrounding slit position. If no $i<22.5$\,mag source was within 10$^{\prime\prime}$ of the slit position (the size of the zCOSMOS long-slit), we matched to a $22.5<i<23.5$ source. In this process we also identified possible repeat observations of the same source (where two slit positions uniquely match to a single source) and zCOSMOS-10k sources which are likely to be secondary sources which fall within the slit of primary target (where two zCOSMOS-10k source uniquely match with a single slit position). We note that in our position matching we do not find a match for 245 zCOSMOS-10k sources. However, none of these sources have unmatched slit positions with 10$^{\prime\prime}$ and as such, are unlikely to be mis-matched to our spectroscopic catalogue. These galaxies are most likely to be targets for which our reduction pipeline has failed.

\begin{figure}
\begin{center}
\includegraphics[scale=0.45]{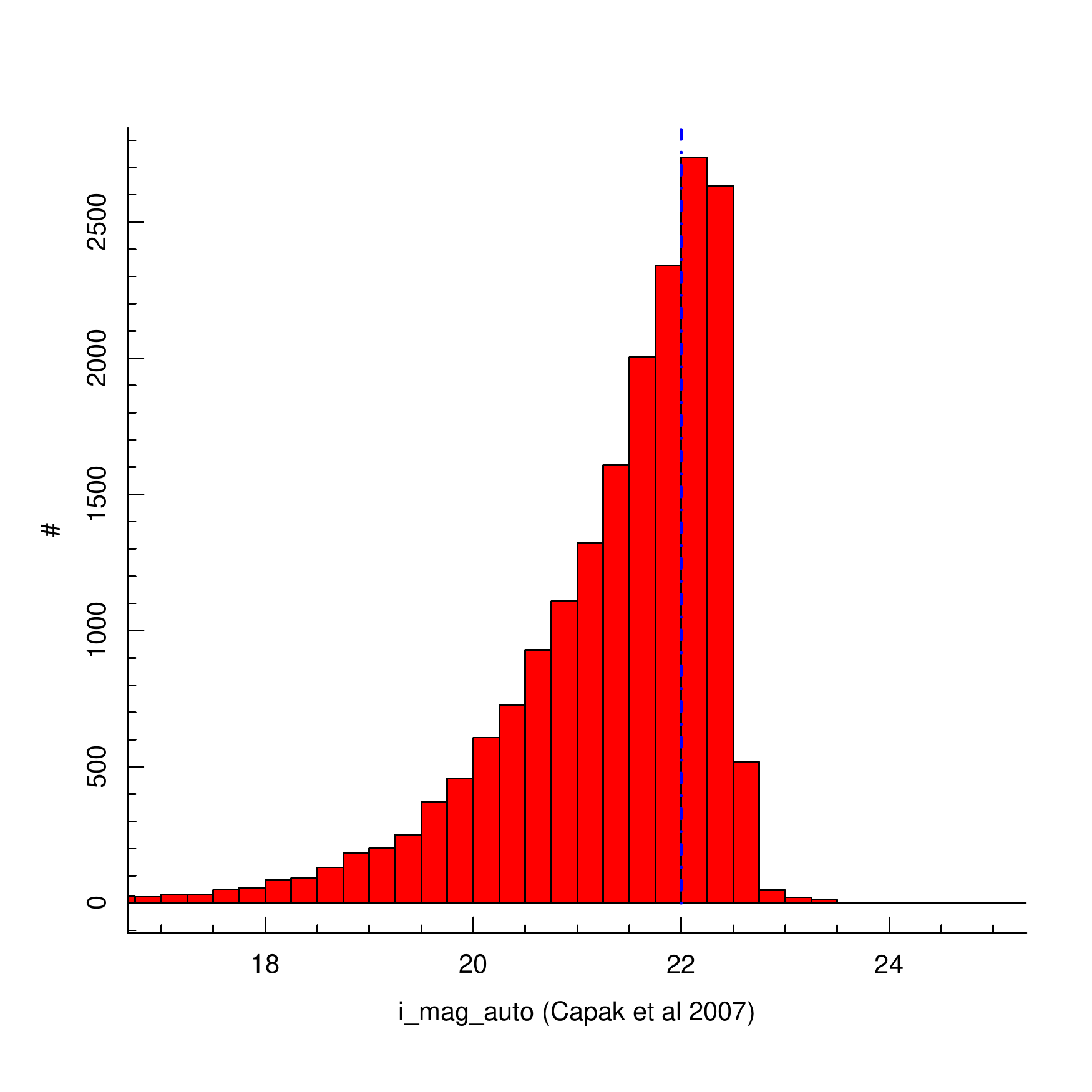}
\includegraphics[scale=0.45]{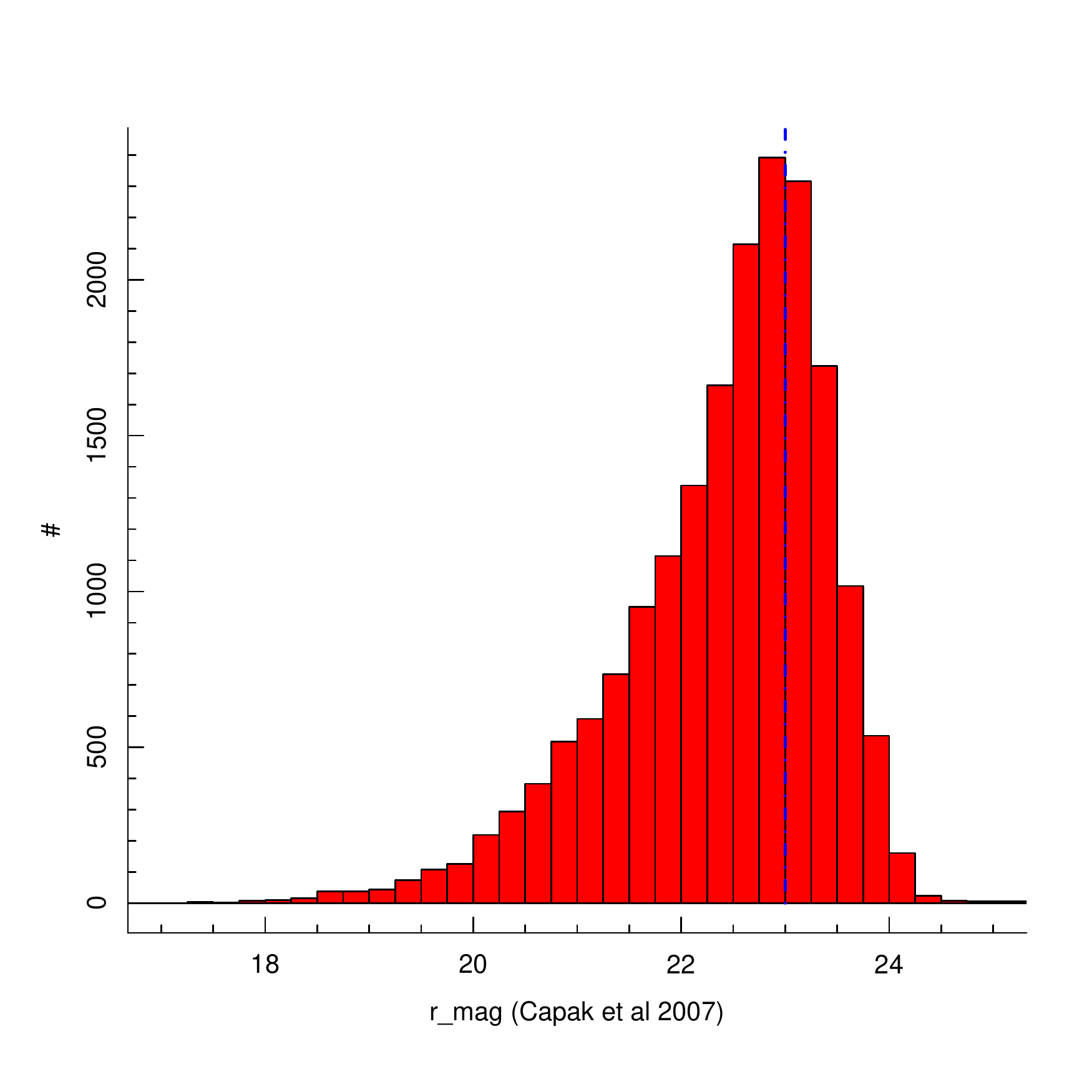}
\caption{The $i$-band (top) and $r$-band (bottom) magnitude distribution of sources matched to our re-reduced zCOSMOS-bright spectra. Blue vertical dashed line displays $i<22.0$ (top) and $r<23.0$ (bottom).}
\label{fig:imag}
\end{center}
\end{figure}

We can estimate the failures in our matching to be consistent with the outlier rate between our redshift distributions (Section \ref{sec:sample}), as any sources which is poorly matched in position are likely to have discrepant redshifts. Comparing our final G10 with zCOSMOS-10k, PRIMUS, and the photometric redshifts, we obtain an outlier rate ($\pm3\sigma$) of $2\%$. Should the zCOSMOS-bright input catalogue be publicly released, it will be possible to test this validity further. The resultant $i$-band and $r$-band magnitude distribution of our re-reduced matched spectroscopic sample can be seen in Figure \ref{fig:imag}. This is consistent with the details of the zCOSMOS-bright input catalogue give in \citep{Lilly07}.

\section{Obtaining the most robust redshift from all available data}
\label{sec:best_red}

In order to obtain the most robust and highest resolution redshifts for the full COSMOS region we compared \textsc{autoz}, zCOSMOS-10k, PRIMUS, VVDS, SDSS and photometric redshifts for all sources in the region. In this manner we automatically assign redshifts as a first pass. In a second step, all sources are visually inspected and automatic redshift reliability assignments (given below) are superseded.    

Firstly, sources were assigned one of the twelve classes given in Table \ref{tab:red_origin}, with a lower class number superseding a higher number from 1-10 (classes 11 and 12 are additionally added VVDS and SDSS spectra). Here we use a liberal matching criterion of redshifts agreeing to within $\pm$10\% of the most robust redshift, but note that all sources are subsequently visually inspected and as such, the 10\% match is only a rough estimate. Classes are: 1=\textsc{autoz} and zCOSMOS-10k redshifts agree to within 10$\%$ (while this data has the same origin it is reduced and redshifted in a completely different manner), 2= \textsc{autoz} and PRIMUS redshifts agree to within 10$\%$, 3=zCOSMOS-10k and PRIMUS agree to within 10$\%$, 4=Reliable zCOSMOS-10k (as indicated by the zCOSMOS team) redshift but does not meet any of the above criteria \cite[quality flags used are the same as those selected by][]{Kovac14}, 5=\textsc{autoz} and photometric redshift agree to within the $1\sigma$ error on the best-fit photometric redshift, 6=robust PRIMUS spectra (as indicated by the PRIMUS team) but no high resolution VIMOS confirmation (Q=3-4 in the PRIMUS catalogues), 7=only \textsc{autoz} redshift with no confirmation from another source, 8=non-reliable zCOSMOS-10k spectrum only, 9=non-reliable PRIMUS spectrum only and 10=no spectroscopic redshift available. In order to maximise the number of redshifts in the COSMOS region, we then additionally added robust VVDS (VVDS team flags Q=2, 3, 4, 22, 23 \& 24) and SDSS (z$_{\mathrm{error}}$ < 0.001 \& CLASS=`GALAXY') redshifts to sources which do not currently have a robust redshift ($i.e.$ classes 5-10). These are assigned classes 11 and 12 respectively. 

For ease of use in our sample we then split our classifications into sources with robust redshifts from VIMOS or SDSS spectra (z\_use=1), robust redshifts from either VIMOS, SDSS or PRIMUS spectra (z\_use=1$\&$2), non-robust redshift but a spectroscopic observation (z\_use=3) and no spectra (z\_use=4). These classifications form the basis of our final G10 sample. Through our automatic redshift reliability assignments we obtain 12,776 robust VIMOS or SDSS redshifts and an additional 5,437 robust PRIMUS redshifts. 

Following our automatic redshift reliability assignments we then visually inspected the 1D and 2D spectra for all $\sim$20,000 zCOSMOS-observed sources and promoted sources with good redshifts to z\_use=1, and demoted those with unclear redshifts to z\_use=3. Our final robust high resolution (z\_use=1) sample contains 16,583 sources in the full COSMOS region. Hence, we gain 3,807 (or $23\%$) of our final sample through our visual classifications.

\begin{table*}
\caption{Breakdown of redshift origins and numbers of reliable redshifts in the full COSMOS region prior to visual classifications, if multiple classes apply, redshift is superseded by a lower class number from 1-10. Notes: $^{a}$Class assigned to object in G10 catalogue (Z\_GEN parameter in final catalogue - see Appendix), $^{b}$number defining whether redshifts should be used in subsequent analysis (Z\_USE parameter in final catalogue): 1=high resolution, reliable redshifts, 2=low resolution redshifts, 3=no reliable redshift, 4=no spectroscopic redshift, $^{c}$number of galaxies in each class prior to visual classifications, $^{d}$redshifts from two surveys agree to within $10\%$, in each case the redshift from the first survey in the name is used, $^{e}$autoz redshift is consistent with the photometric redshift within the $1\sigma$ error given in the \citet{Ilbert09} catalogue,  $^{f}$only \textsc{autoz} spectroscopic redshift available - note all sources were subsequently visually inspected and a large fraction were added to the final catalogue. }

\begin{center}
\begin{tabular}{c c c c }
Class$^{a}$  & Origin & z\_use$^{b}$ & $\#$$^{c}$  \\ 
\hline 
\hline 
1 & \textsc{autoz}-zCOS$^{d}$ & 1 & 3,878, \\
2 & \textsc{autoz}-PRIMUS$^{d}$& 1 & 2,024 \\
3 & zCOS-PRIMUS$^{d}$ & 1 & 2,320   \\
4 & zCOS robust & 1 & 3,130  \\
5 & \textsc{autoz }-zPhoto$^{e}$ & 1 & 1,424 \\
6 & PRIMUS Q=3 | 4 & 2 & 6,768 \\
7 & \textsc{autoz} $^{f}$ & 3 & 5,434 \\
8 & zCOS not-robust & 3 & 225 \\
9 & PRIMUS Q$\neq$3 | 4 & 3 & 14,116 \\
10 & zPhoto & 4 & 398,200 \\

\hline 
11 & VVDS & 1 & 340 \\
12 & SDSS-DR10 & 1 & 367 \\

\hline 
 &Total HR pre-visual class (z\_use=1) & & 12,776\\
 &Total HR after visual class (z\_use=1) & & 16,583\\
 
\hline 
 
 &Total ALL pre-visual class (z\_use<=2) & & 19,544 \\
 &Total ALL after-visual class (z\_use<=2) & & 22,020 \\
\hline 
\end{tabular}
\end{center}
\label{tab:red_origin}
\end{table*}

\begin{figure*}
\begin{center}
\includegraphics[scale=0.3]{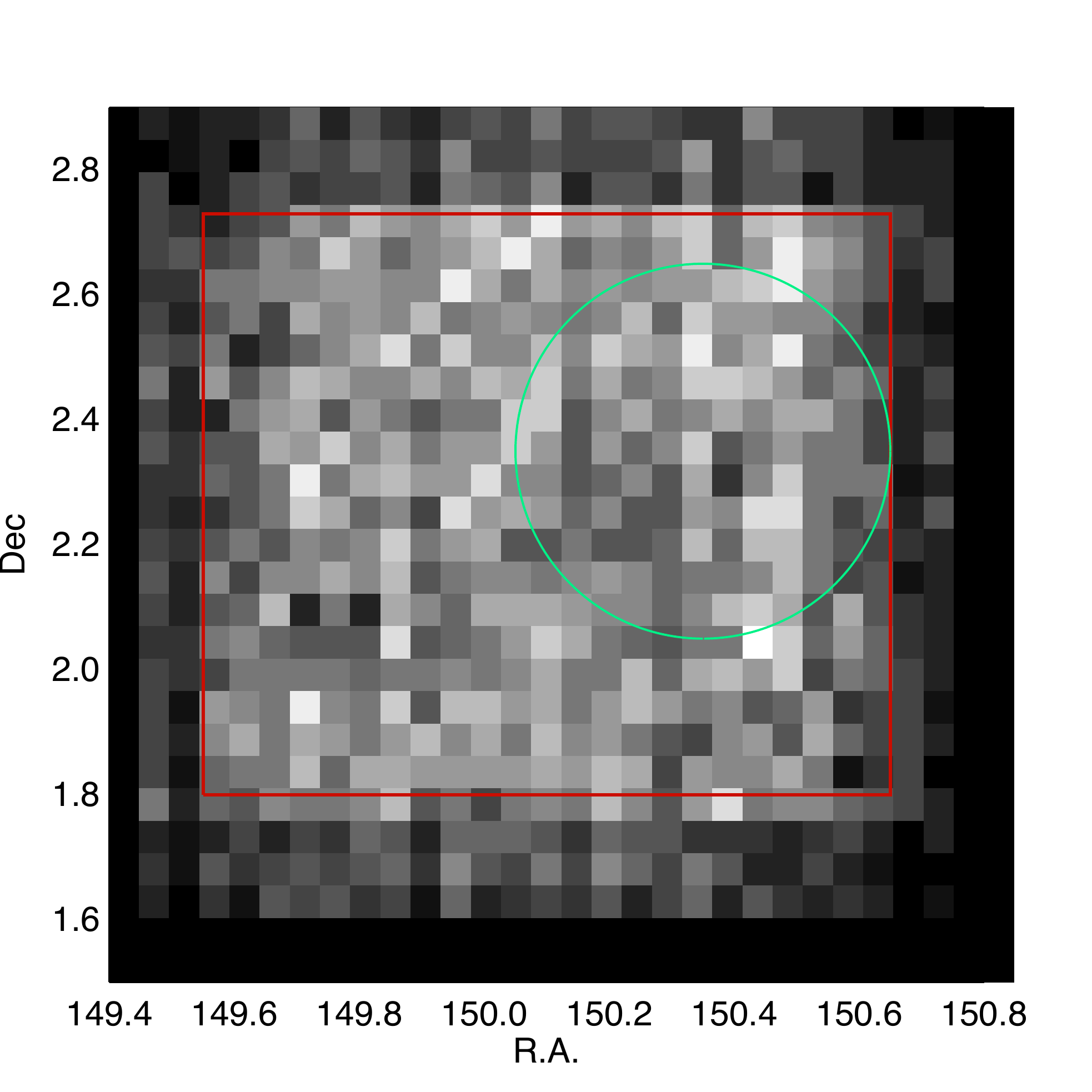}
\includegraphics[scale=0.3]{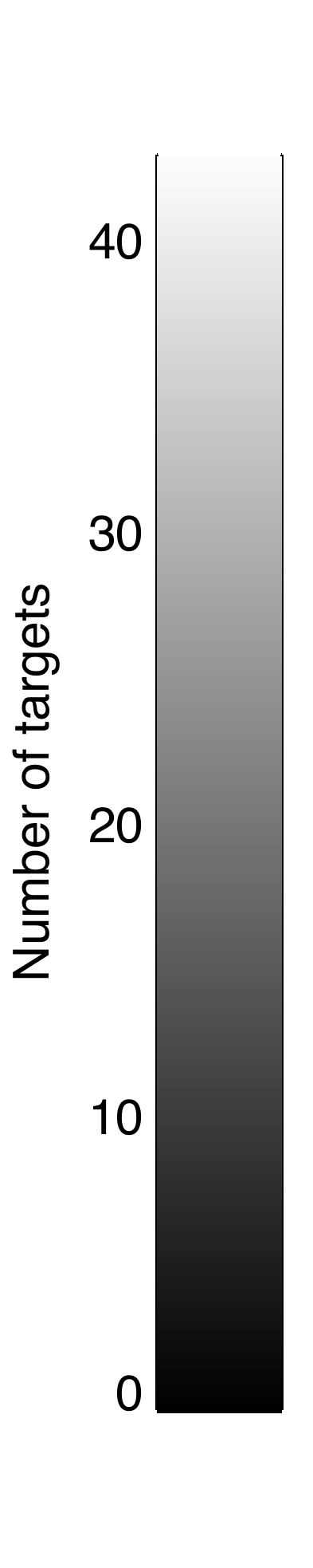}
\includegraphics[scale=0.3]{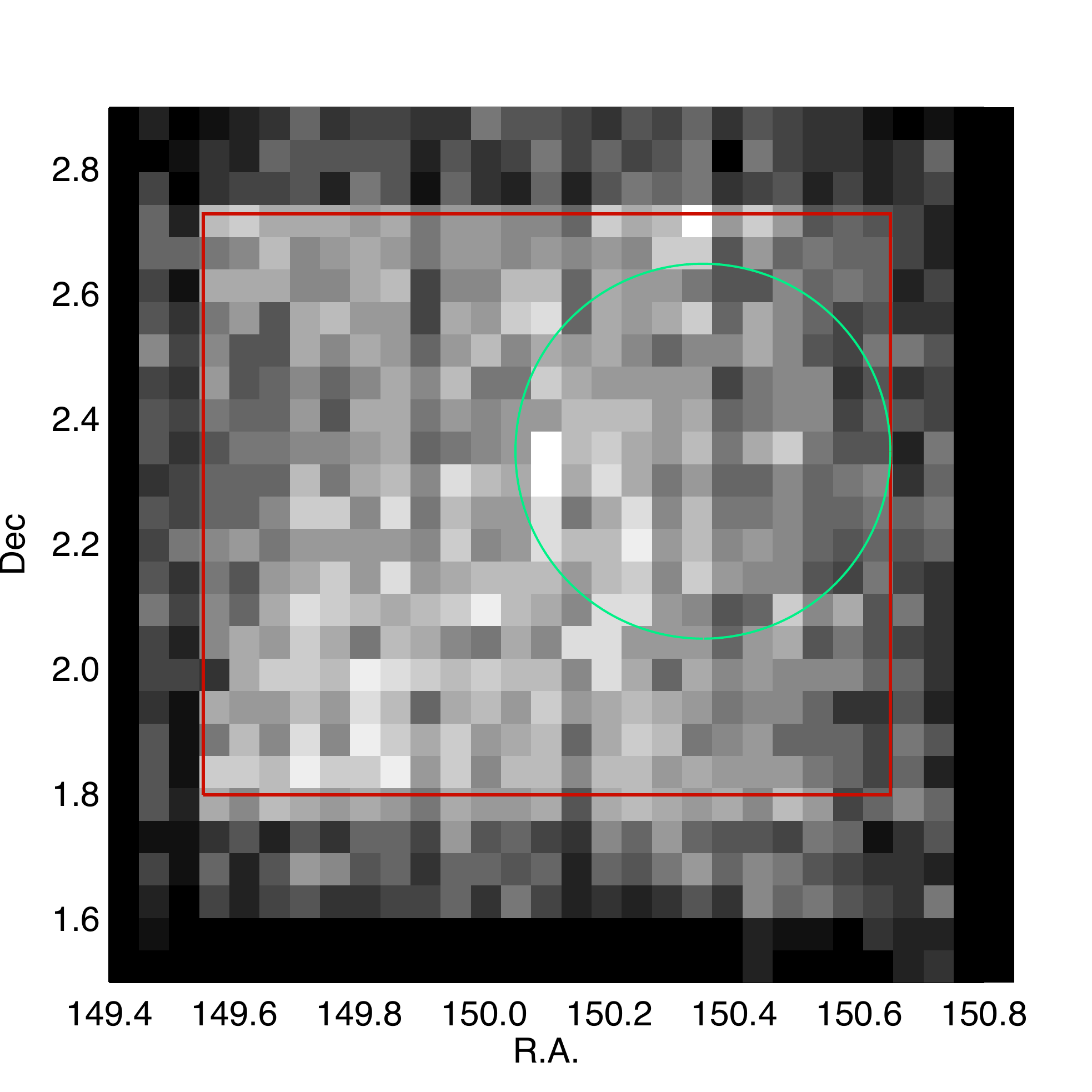}
\includegraphics[scale=0.3]{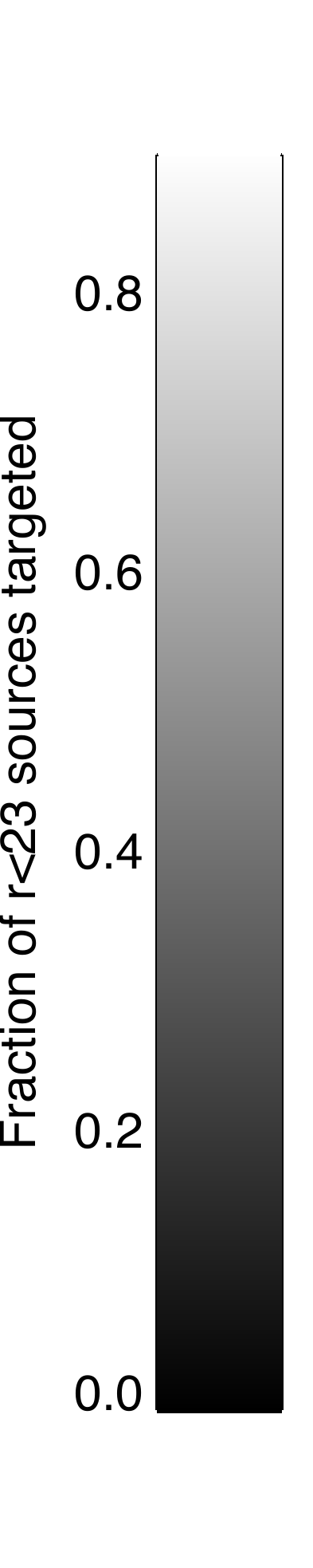}\\

\includegraphics[scale=0.3]{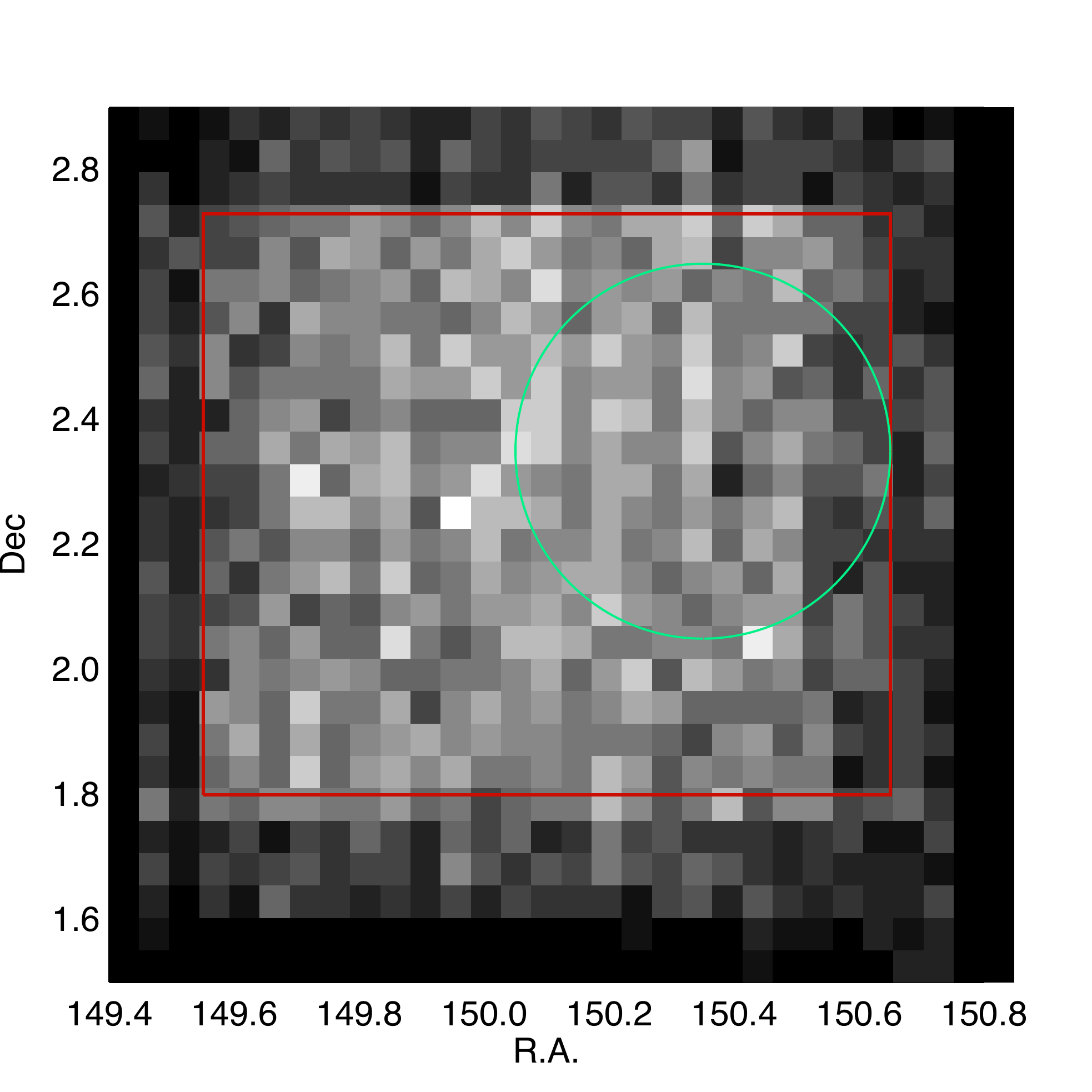}
\includegraphics[scale=0.3]{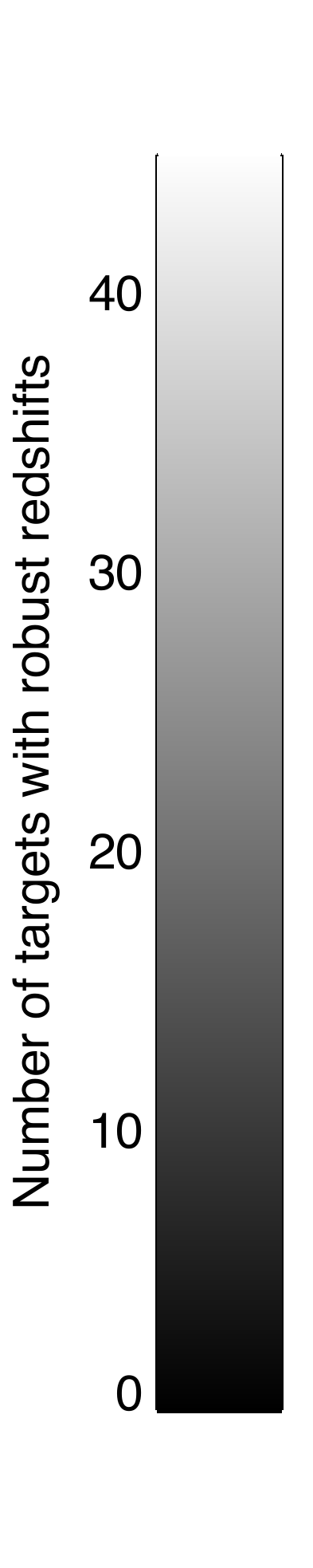} 
\includegraphics[scale=0.3]{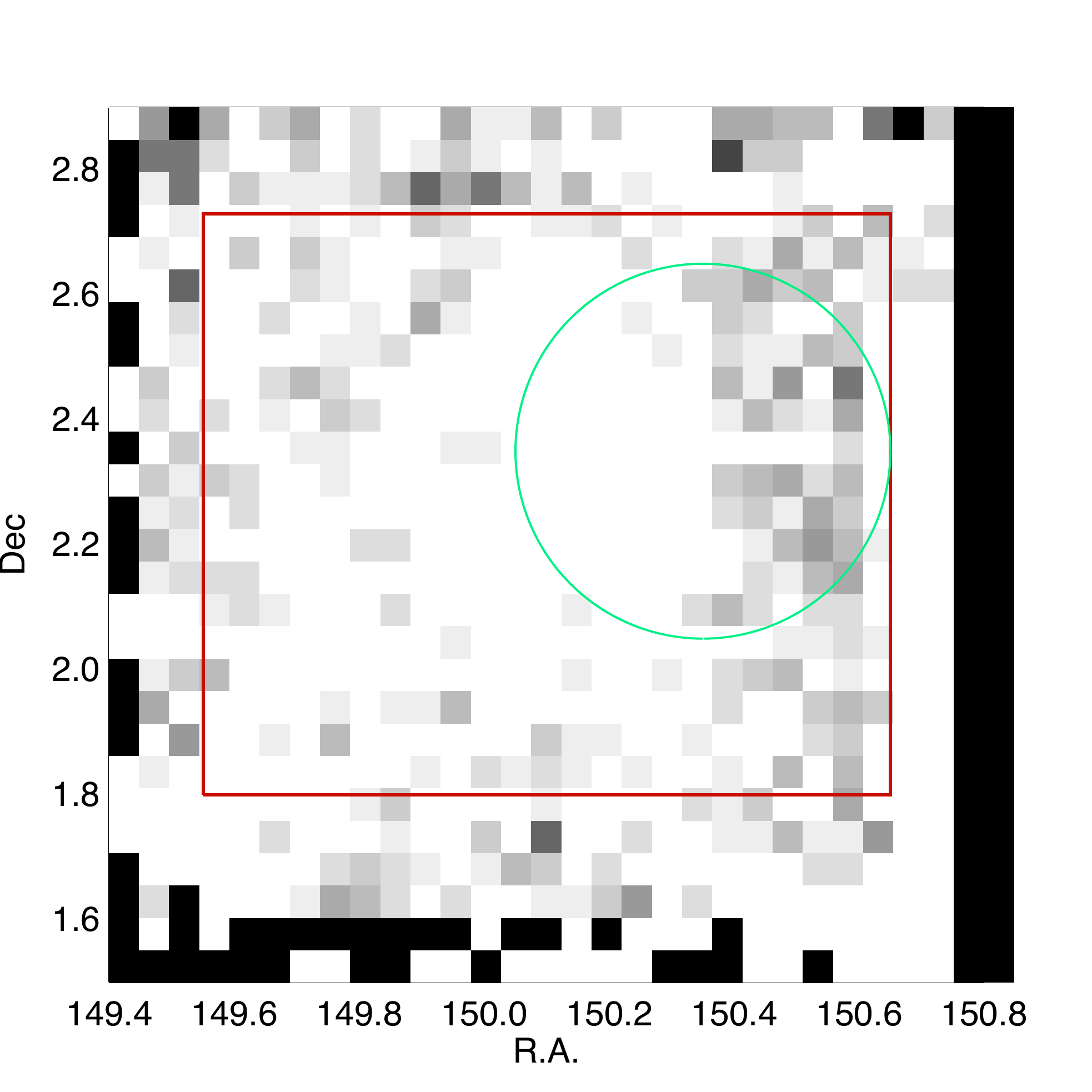}
\includegraphics[scale=0.3]{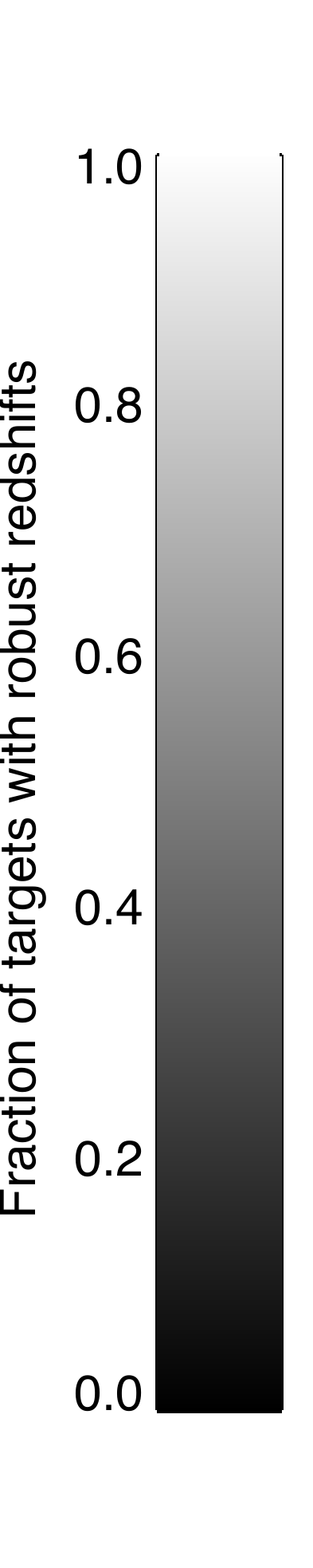}

\caption{zCOSMOS-bright spectroscopic coverage for $r<23.0$\,mag \& $i<22.0$\,mag combined  sources in the COSMOS region binned on 3$^{\prime}$ scales, light colours represent high number/fraction of sources. Top left: density of zCOSMOS spectroscopic targets, top right: fraction of all $r<23.0$\,mag \& $i<22.0$\,mag combined sources targeted by the zCOSMOS observations ($i.e.$ the target sampling rate), bottom left: number of targets with robust spectroscopic redshifts in our full sample and bottom right: fraction of zCOSMOS observed sources with robust spectra in our full sample. We define the G10 region as the area bounded by the red box, this region has high completeness and includes the CHILES VLA region (cyan circle).}
\label{fig:good_density}
\end{center}
\end{figure*}

\begin{figure}
\begin{center}
\includegraphics[scale=0.45]{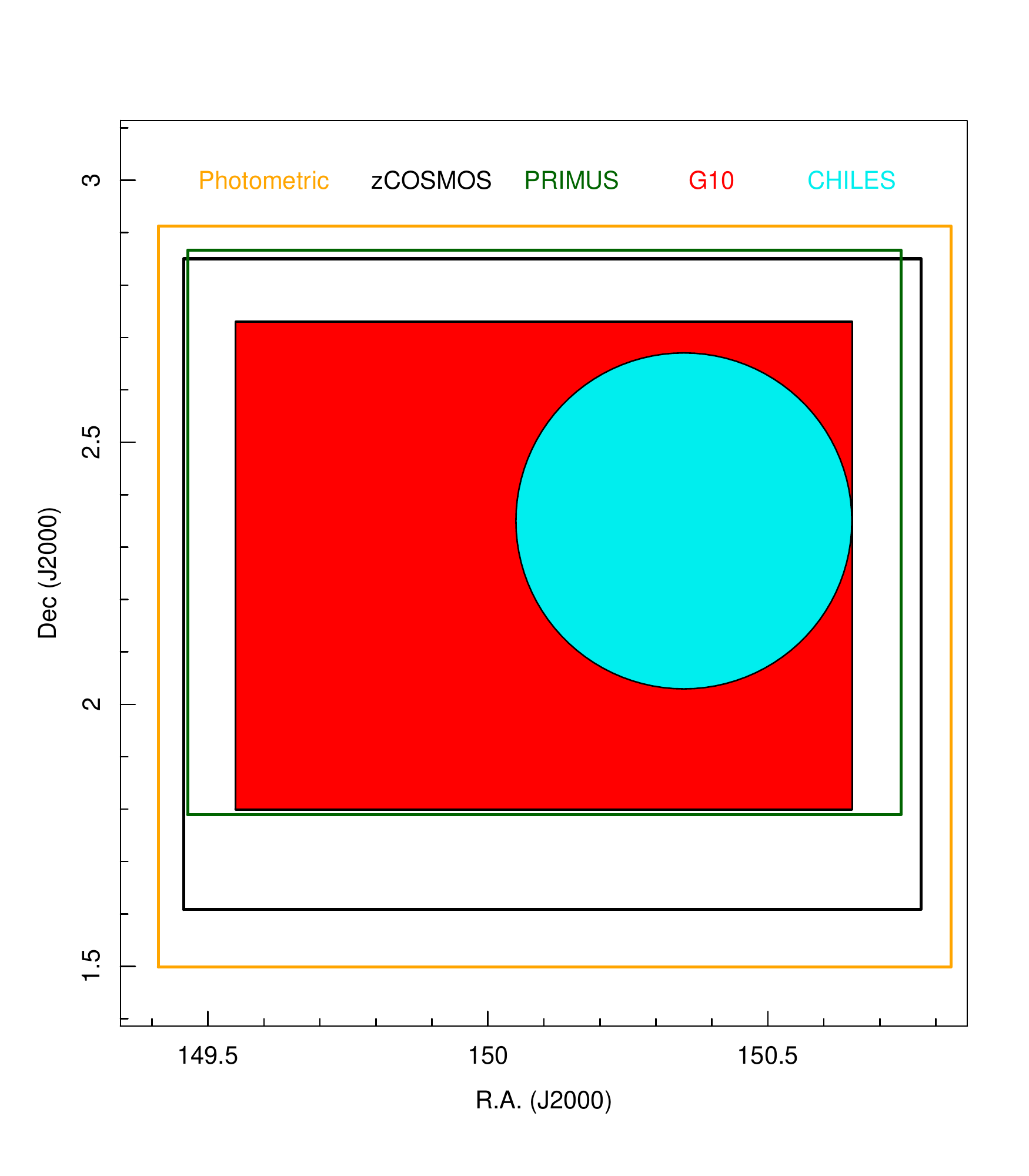}

\caption{The G10 region in comparison to other redshift surveys in the COSMOS region. The Photometric region describes the deep photometric data available in the COSMOS region, including the HST observations. The black region displays the full extent of zCOSMOS observations, while the red box displays the G10 region covering the central, high completeness region of zCOSMOS and cyan circle highlights the CHILES region.}
\label{fig:regions}
\end{center}
\end{figure}

\begin{figure*}
\begin{center}

\includegraphics[scale=0.45]{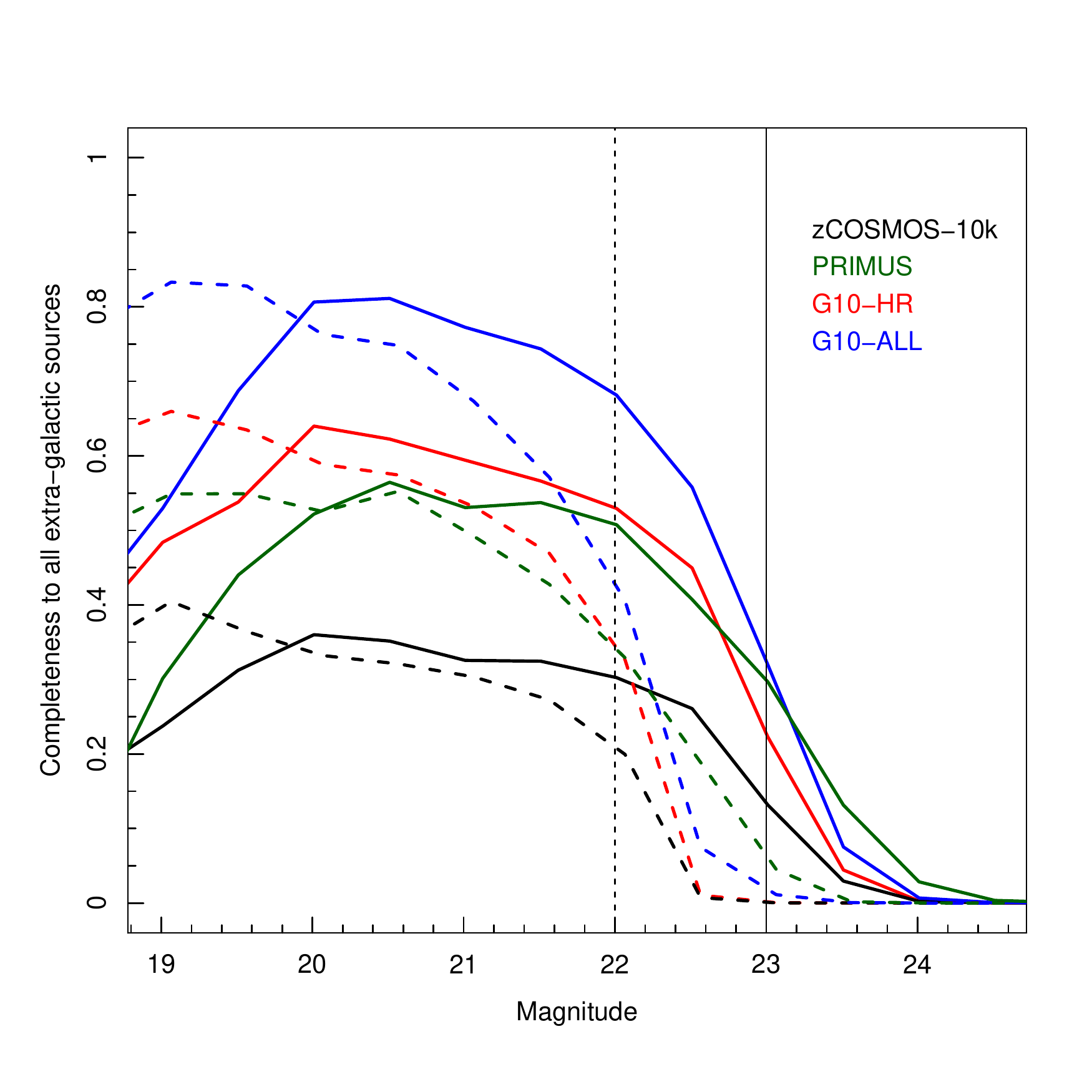}
\includegraphics[scale=0.45]{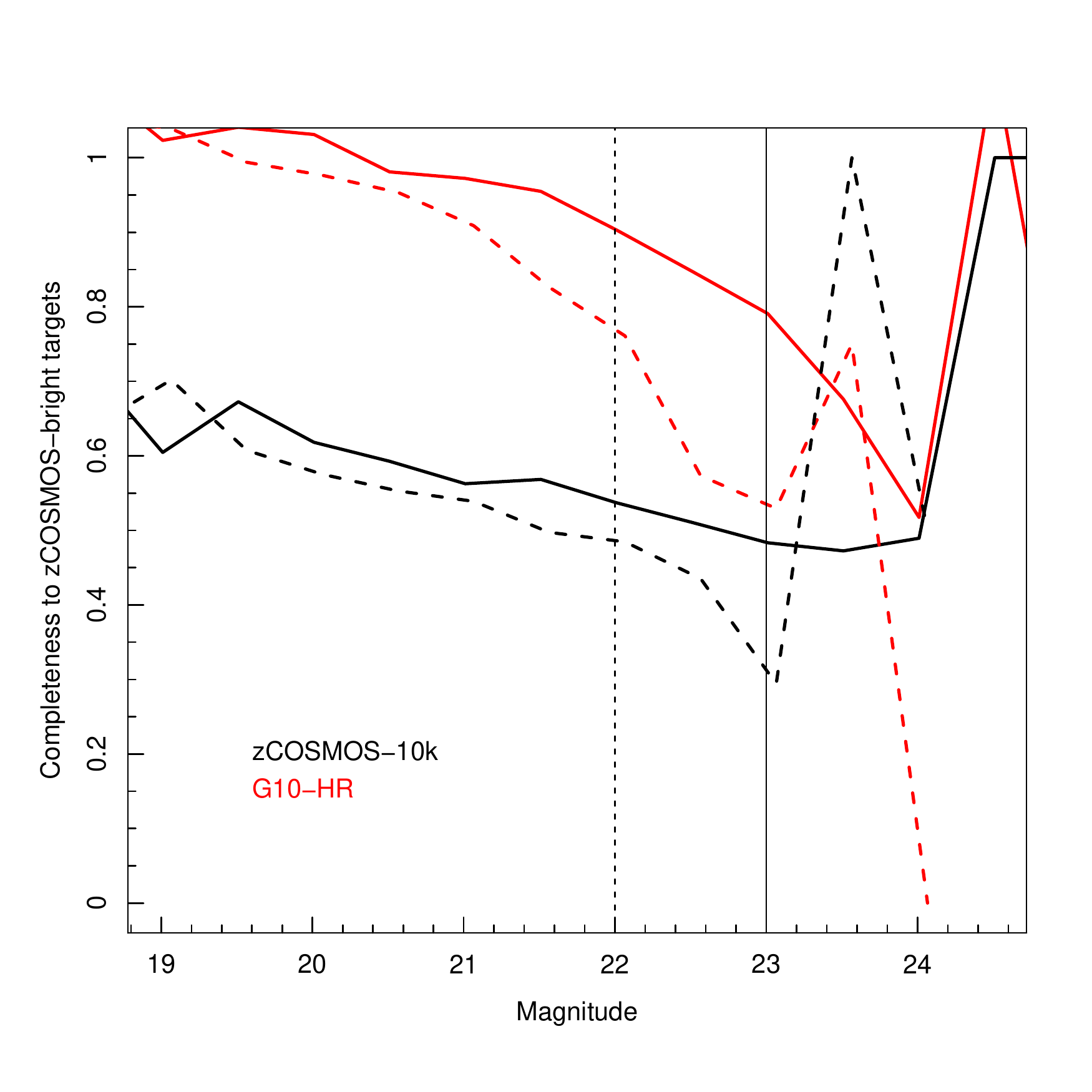}

\caption{Spectroscopic completeness for the G10 region in both $r-$band (solid lines) and $i$-band (dashed lines). Left: The completeness with respect to all galaxies in the field, right: the completeness of zCOSMOS-bright spectroscopic targets in the G10 region. Black line displays zCOSMOS and red line displays the G10-HR completenesses. At $r>24$ \& $i>23$ the completeness is erratic as only a small number of sources are targeted. }
\label{fig:completeness}
\end{center}
\end{figure*}

\section{The G10 sample}
\label{sec:sample}

To define the G10 region we considered the target density, and spectroscopic success rate of the full zCOSMOS-bright survey taken from our sample of high resolution, robust spectra outlined above. Figure \ref{fig:good_density} displays the total number (left column) and fraction (right column) of possible photometric targets at $r<23.0$\,mag \& $i<22.0$\,mag combined (top) and robust spectroscopic redshifts (bottom) - see below for magnitude choices. Hence, the top panels show the source completeness of zCOSMOS-bright for all extragalactic sources at $r<23.0$\,mag \& $i<22.0$\,mag, and the bottom panels display the spectroscopic success rate when using our robust catalogue. From the top right panel it is clear that our final robust sample has a high completeness in the central region of the zCOSMOS observations. We defined the G10 region as the red box bounding the region of both high target density, good spectroscopic completeness and to fully encompass the CHILES survey region (cyan circle). Figure \ref{fig:regions} shows the position of the G10 region with respect to the full zCOSMOS region, PRIMUS and photometric observations used in this work as well as the CHILES region. The photometric region is essentially the area covered by the deep photometric observations on COSMOS, including the HST data. The G10 region covers a $\sim$1\,deg$^{2}$ region with R.A.=149.55 - 150.65, $\delta$=1.80 - 2.73, and represents 0.55 of the total COSMOS area. In our subsequent description of the sample in this work we only include sources within this G10 region. However, our publicly available catalogues contain the most robust redshift estimates for all sources in the full COSMOS region.

Considering just the G10 region, we have defined two samples with which subsequent GAMA-like analysis will be performed. Firstly, we select all sources with robust high precision spectroscopic redshifts (z\_use=1, Table \ref{tab:red_origin}) in the G10 region as the G10-HR sample, and sources with either a robust high precision redshift or a robust low precision PRIMUS redshift as the G10-ALL sample (z\_use=1 and z\_use=2, Table \ref{tab:red_origin}). The left panel of Figure \ref{fig:completeness} shows the completeness of each sample as a function of $r$-band (solid lines) and $i$-band (dashed lines) magnitude, in comparison to robust spectroscopic redshifts from the zCOSMOS-10k sample and the PRIMUS catalogues. The right panel shows the completeness of spectroscopic targets observed in zCOSMOS-bright for our G10-HR sample and the zCOSMOS-10k release. 

To compare our sample to extensive low redshift surveys (like GAMA), we define it to be magnitude limited. However, selecting an identical sample to those defined in the local Universe is not straightforward, as galaxies would ideally be included on rest-frame magnitudes. At low-$z$, GAMA selects galaxies on observed $r$-band magnitude ($r<19.8$), and while a single band selection has little bias on the true rest-frame magnitudes over the epochs probed by GAMA, an $r-$band selection becomes problematic at $z>0.4$ when the $4000$\AA\ spectral feature begins to play a significant role in the observed $r$-band flux (where the bulk of our G10 sample lies). A possible solution is to use $i$-band selections to probe similar rest-frame fluxes to the $r$-band at low redshift (such as zCOSMOS-bright). However, these samples then have complex selection functions when equating to low redshift surveys as one can not compare rest-frame flux selected samples over a large redshift baseline. Here we define both an $r$- and $i$-band magnitude limited sample simultaneously, selecting our sample limits to cover the largest bijective population, and then allow future studies to post engineer a more sophisticated sample selection. In this manner a sample can either be defined in the $r$-band, with which to compare directly to low-$z$ systems but pushing to fainter magnitudes, in the $i$-band, to target similar rest-frame magnitudes at high redshift, or even using a sliding $r-i$ colour, to target true rest-frame selected samples and compare galaxies over a large redshift baseline.

\begin{figure}
\begin{center}
\includegraphics[scale=0.45]{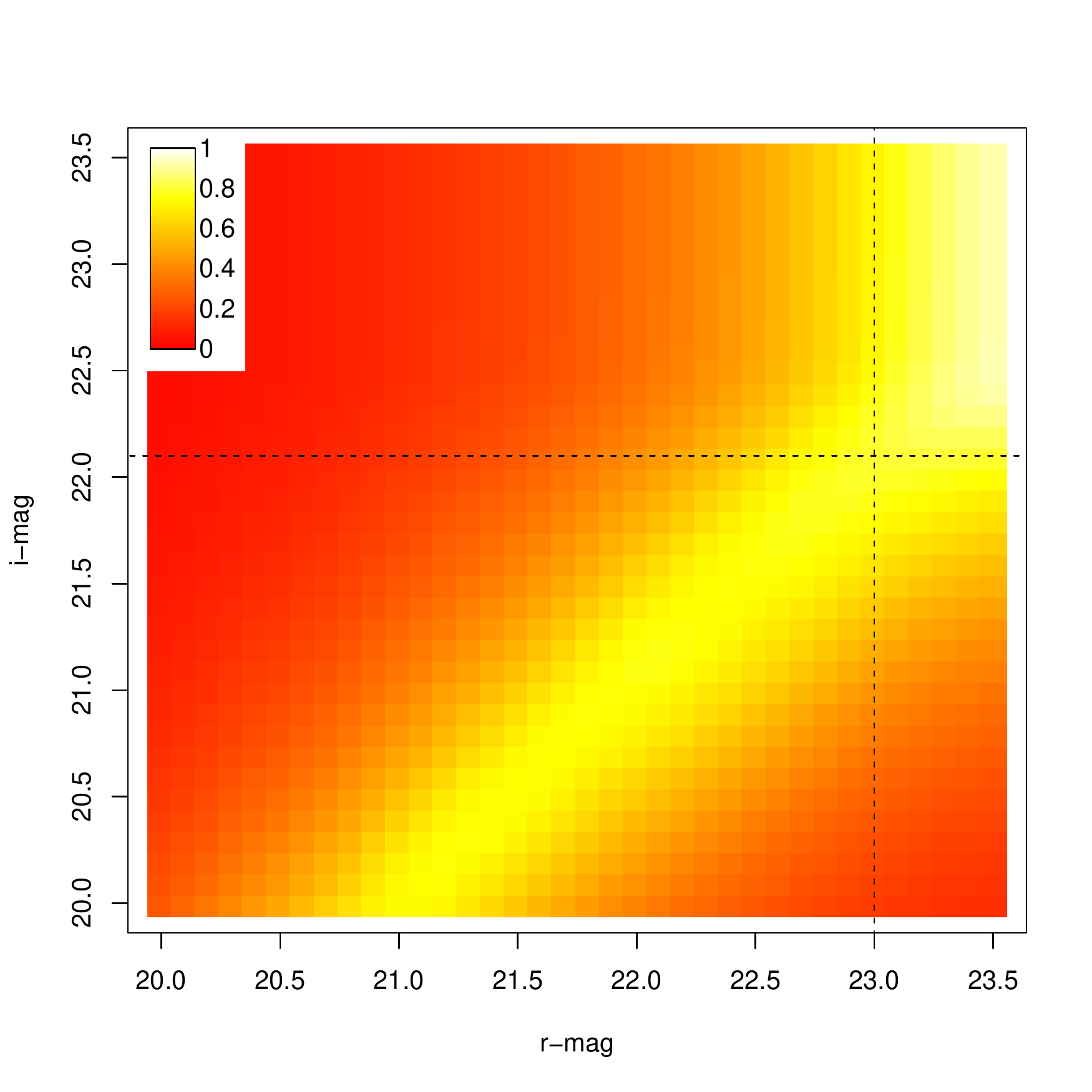}

\caption{The bivariant distribution of $r-$ and $i-$band magnitudes in the G10-HR sample. An $r-$band magnitude limit of $r<23$\,mag has the largest bijective overlap with an $i-$band limit of $i<21.1$\,mag. }
\label{fig:bijective}
\end{center}
\end{figure}

We define the $r$-band magnitude limit from our G10 samples at $r<23$\,mag (solid vertical line in Figure \ref{fig:completeness}) prior to the rapid decrease in spectroscopic completeness. To define a corresponding $i$-band limit, we consider the bivariant distribution of $r$-band and $i$-band magnitudes in our G10-HR sample (Figure \ref{fig:bijective}). At $r=23$, the largest bijective overlap in populations occurs at $i=22.1$ (the $i$-mag value which have the largest sample overlap with an $r<23.0$ population - dotted lines), for sake of clarity in our sample we scale this limit to $i<22.0$ (dashed vertical line in Figure \ref{fig:completeness}).

Our final G10 samples are defined as sources with robust redshifts falling within the G10 region and being below an $r<23$\,mag OR $i<22$\,mag limit. The number and completeness of sources in these samples, and for the full COSMOS region are given in Table \ref{tab:compare}. Clearly our samples are much improved over the zCOSMOS-10k data release. We increase the number of high resolution robust redshifts in the G10 region at $r<23$\,mag or $i<23$\,mag from 5,670 to 9,861 while obtaining a higher sample spectroscopic completeness of zCOSMOS targets. It is interesting to note that we also obtain a higher spectroscopic completeness than the PRIMUS survey at an order of magnitude better spectral resolution (150\,km\,s$^{-1}$ compared to 1500\,km\,s$^{-1}$). In the following we display all samples with the $r<23$\,mag \& $i<23$\,mag cut applied.

\begin{figure}
\begin{center}
\includegraphics[scale=0.5]{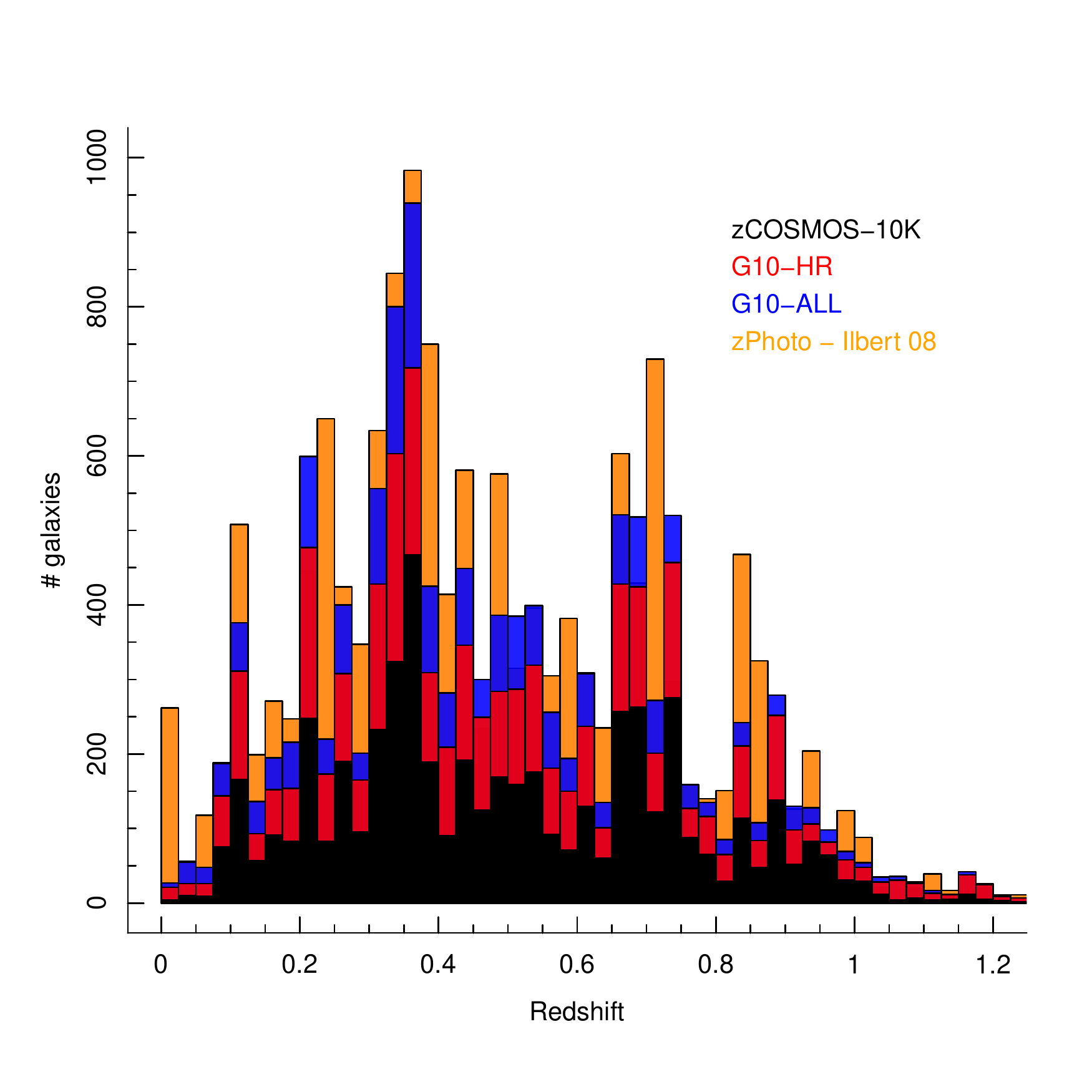}

\caption{The redshift distribution of $r<23$\,mag \& $i<22$\,mag galaxies in the G10 region for our G10-HR sample (red), G10-ALL (blue) and the zCOSMOS-bright 10k sample (black) in comparison to the photometric redshift distribution (all galaxies at $r<23$ \& $i<22$\,mag, orange). Note the the histogram bars are $not$ cumulative ($i.e.$ the top of each coloured bar shows the total number at each redshift). It is interesting to note that we improve upon the zCOSMOS-bright 10k sample at all redshifts. }
\label{fig:red_dist}
\end{center}
\end{figure}

\begin{figure*}
\begin{center}
\includegraphics[scale=0.6]{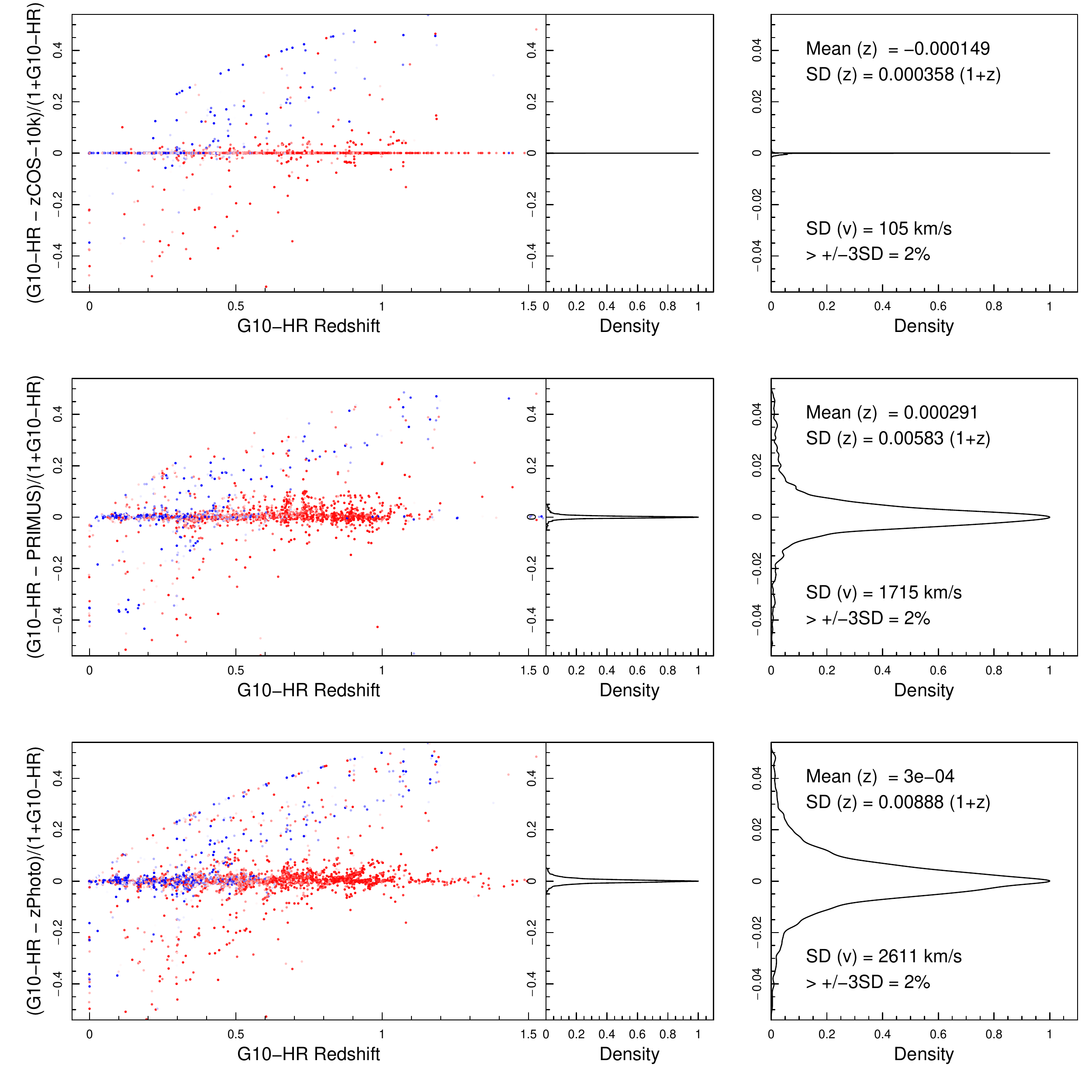}

\caption{Comparison between our G10-HR sample and various other redshift campaigns in the region. Left displays the offset between our G10 redshifts and previous surveys as a function of redshift ($y=\Delta z/[1+z]$), with $\Delta z$ density. Right shows a zoomed in region of the $\Delta z$ density surrounding the peak. Sources are coloured by apparent $r$-band magnitude ranging from bright (darkest blue) to faint (darkest red). 3$\sigma$ clipped means and standard deviations are given in both redshift and velocity space. We also display the percentage of outliers at $\Delta \sigma > \pm3\sigma$ for each sample.}
\label{fig:red_comp}
\end{center}
\end{figure*}

Figure \ref{fig:red_dist} shows the redshift distributions of our samples in comparison to the zCOSMOS-10k release and the photometric redshifts of \cite{Ilbert09}. We find an improvement over the zCOSMOS-10k release at all redshifts and also note that our final redshift distribution is similar to both the photometric redshifts and zCOSMOS, suggesting there is likely to be no redshift biases in our analysis. Comparing the redshifts from each sample directly, Figure \ref{fig:red_comp} shows the redshift offset between our G10-HR sample and the previous redshift surveys in the COSMOS region. The colour coding in this figure displays apparent $r$-band magnitude (from blue=bright to red=faint). There is some degeneracy in the top panel of this figure, as a number of the redshifts in the G10-HR sample are derived from the zCOSMOS-10k catalogue directly. However, this figure displays that our subsequent analysis does not deviate largely from the previous zCOSMOS results. In each Figure we highlight the $3\sigma$-clipped mean and standard deviations of the offsets between each sample in both redshift and velocity.  This suggests that there is little or no systematic error in our redshift analysis and the redshifts derived in our G10 sample are consistent with previous studies.      

In order to visualise the 3D distribution of galaxies in our G10-HR sample we produce light cones for all sources out to $z\sim1.2$ (Figure \ref{fig:light_cones} - once again coloured according to $r$-band apparent magnitude). The panels are split into $\Delta z=0.15$ sections in order to highlight large scale structure. We find similar large scale structure to that found in \cite{Knobel12} who perform a group finding analysis over the zCOSMOS data (see their Figure 7), such as the over-densities of sources at $z\sim0.22$ and at $0.34<z<0.38$. The light cone distribution displayed in Figure \ref{fig:light_cones} will form the basis of our own group finding in the G10 region performed in both a consistent  manner to the GAMA analysis \citep[$e.g.$][]{Robotham11} and using newly developed group identification techniques (Kafle et al, in prep).

\begin{table*}
\caption{Comparison to previous surveys in the full COSMOS and G10 regions (with magnitude limits applied). The last column displays the completeness of each survey to all extra-galactic $r<23$\,mag \& $i<22$\,mag sources in the G10 region.}
\begin{scriptsize}
\begin{center}

\begin{tabular}{c c c c c c c c}
Survey & Total Targets & Robust redshifts & $\%$ robust& Total Targets  & Robust redshifts  & $\%$ robust& $\%$ robust\\ 
            & (all) & (all) & (all targets) &  (G10-mag) & (G10-mag) & (G10-mag) & (all G10-mag) \\
           
\hline 
\hline 

HR & 19,120 & 16,583 & 87$\%$ & 11,188 & 9,861 & 88$\%$ & 51$\%$ \\
ALL & 38,947 & 22,020  & 57$\%$ & 16,624 & 12,027 & 72$\%$ & 65$\%$\\
zCOSMOS-10k & 10,632  & 8,999 &  86$\%$ & 6,515 & 5,670 & 87$\%$& 29$\%$\\
PRIMUS-COS & 29,312 &  17,649 & 60$\%$ & 12,027 & 8,902 & 75$\%$& 46$\%$\\

\hline 

\end{tabular}

\end{center}
\end{scriptsize}
\label{tab:compare}
\end{table*}

\begin{figure*}
\begin{center}

\includegraphics[scale=0.45]{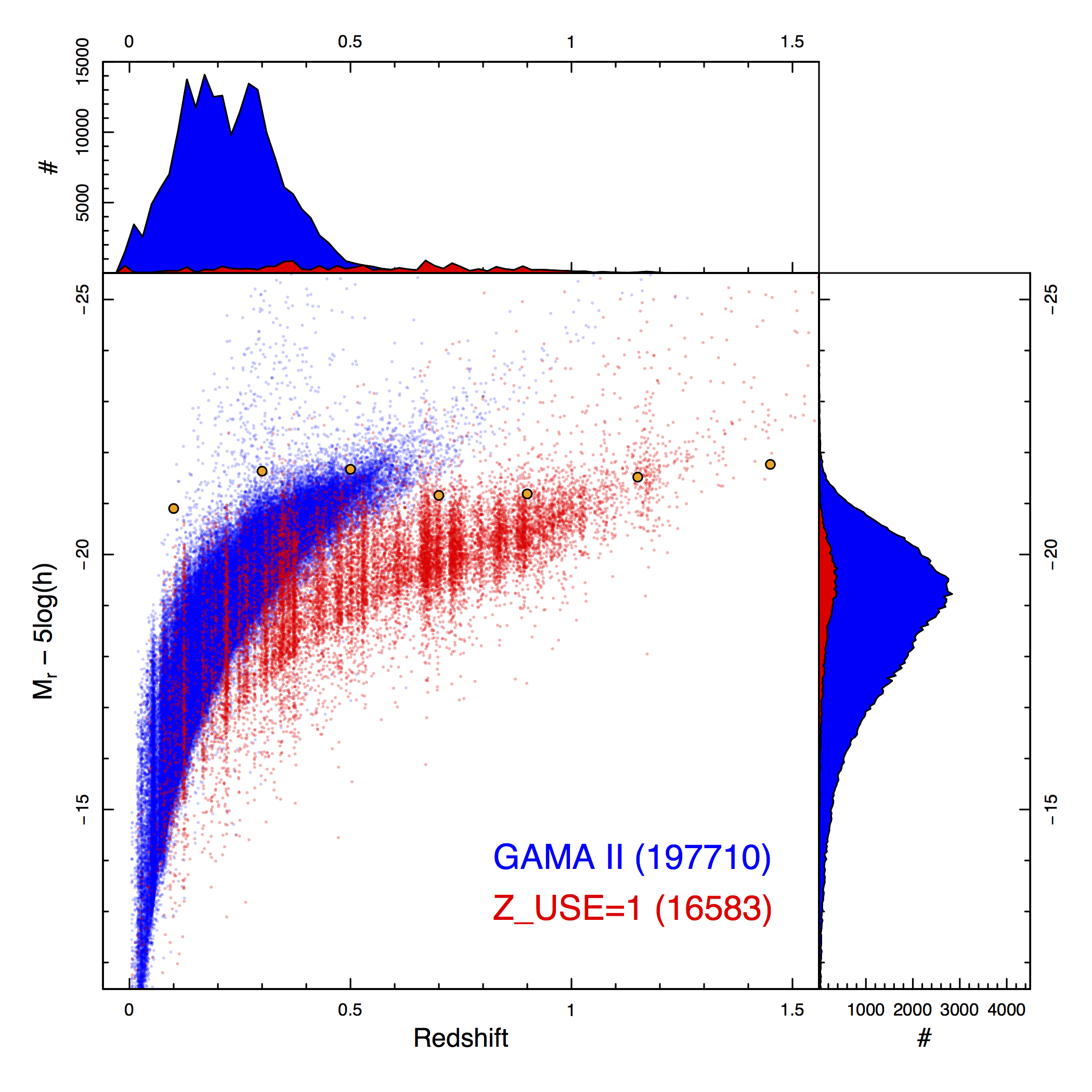}
\includegraphics[scale=0.45]{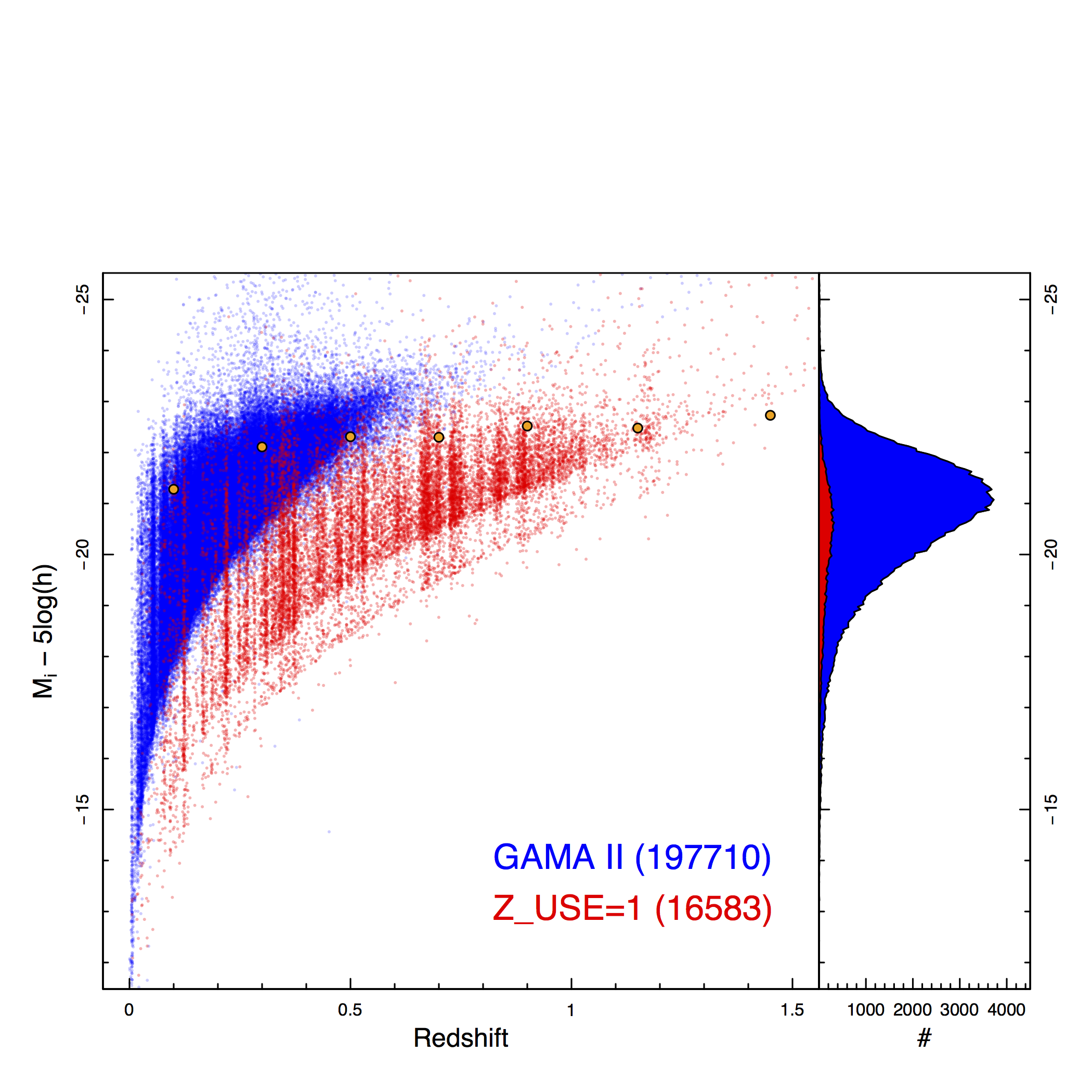}

\caption{Main panels display the absolute magnitude versus redshift for
the G10 region in comparison to the GAMA II equatorial regions ($r$-band left, $i$-band right). Top panel show
the distribution collapsed in redshift, while the right panels show the distribution
collapsed in absolute magnitude. The orange circles display M$_{i}^{*}$ as a function of redshift taken from \citet{Ramos11}. In the left panel these are scaled to M$_{r}^{*}$ using the median $r-i$ colour at $\Delta z\pm0.1$ in a combined GAMA II+G10 sample.}
\label{fig:Mr_dist}
\end{center}
\end{figure*}

\begin{figure*}
\begin{center}

\vspace{-6mm}

\includegraphics[scale=0.55]{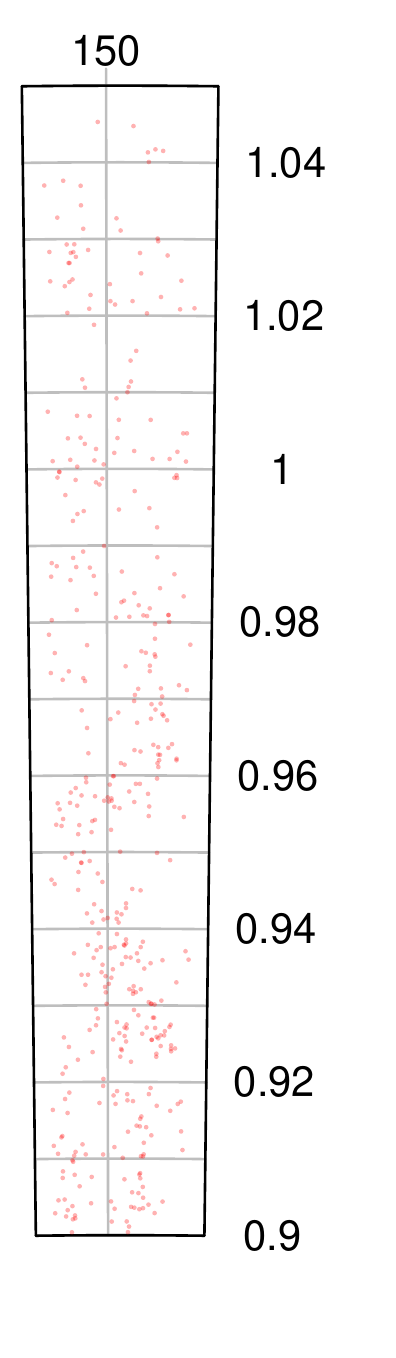}
\hspace{10mm}
\includegraphics[scale=0.55]{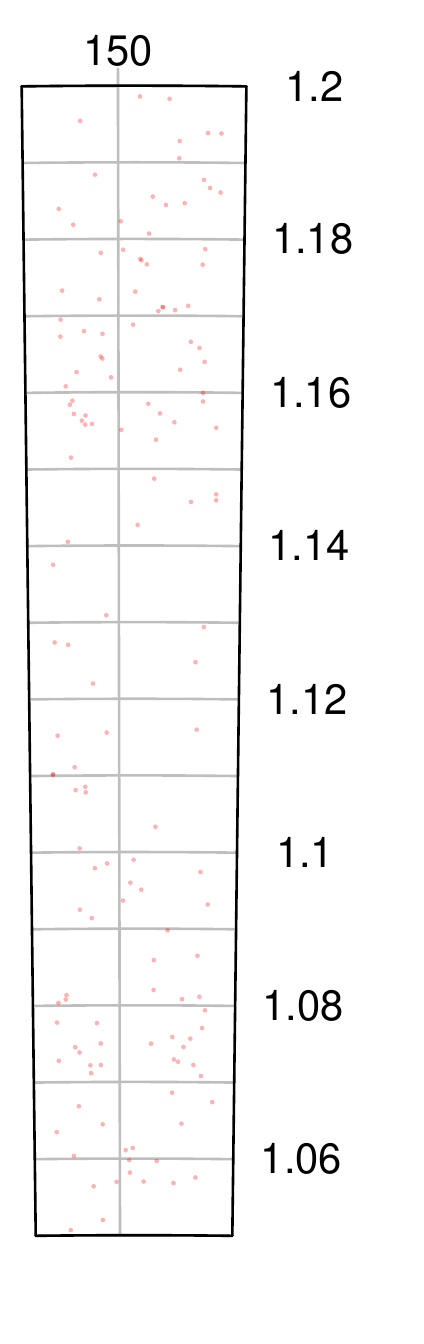}
\hspace{14mm}
\vline
\hspace{14mm}
\includegraphics[scale=0.55]{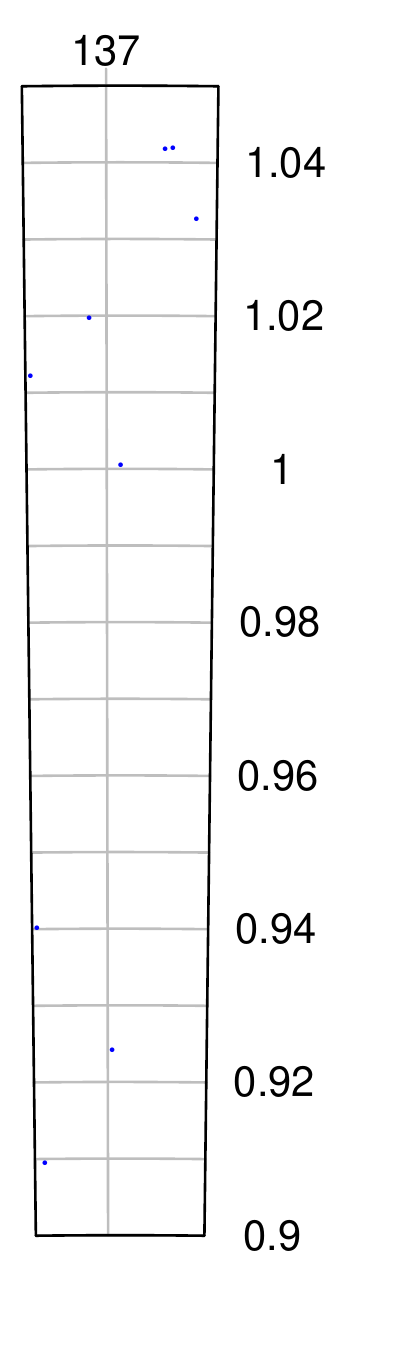}
\hspace{10mm}
\includegraphics[scale=0.55]{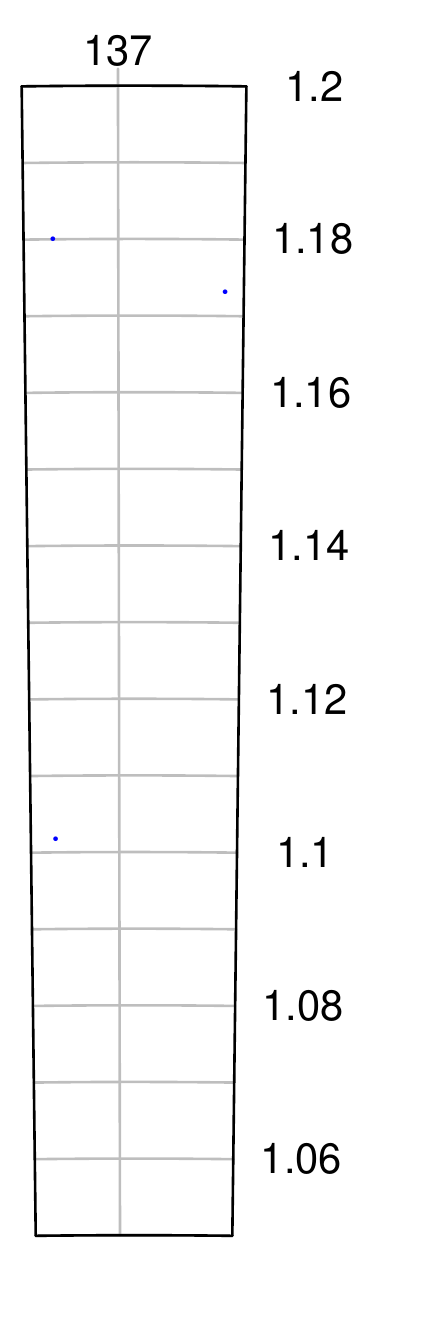}\\

\vspace{-6mm}

\includegraphics[scale=0.6]{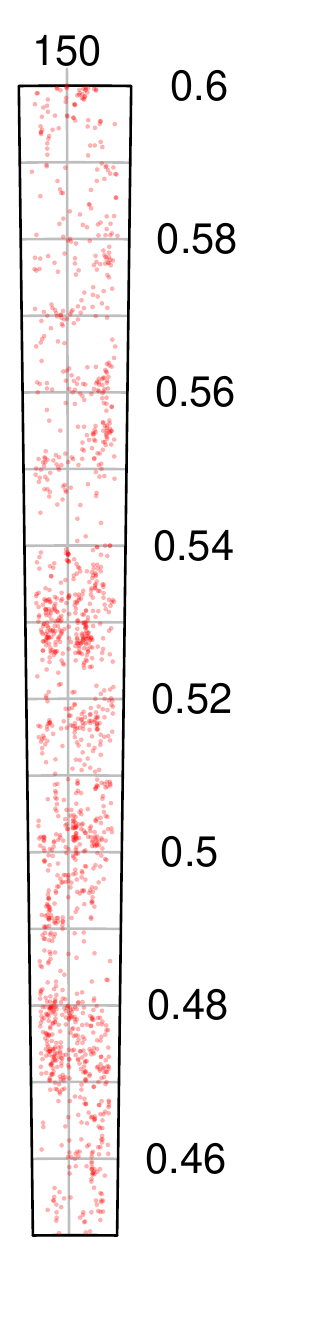}
\hspace{6mm}
\includegraphics[scale=0.6]{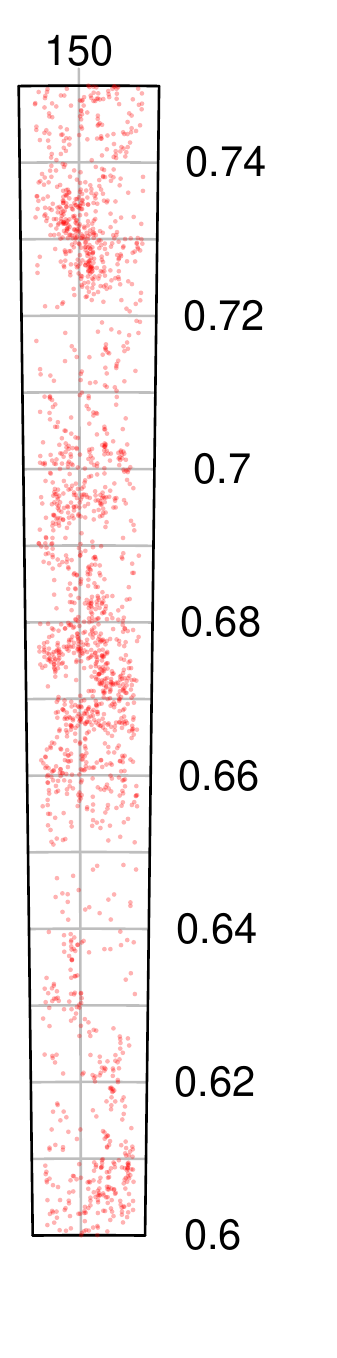}
\hspace{6mm}
\includegraphics[scale=0.6]{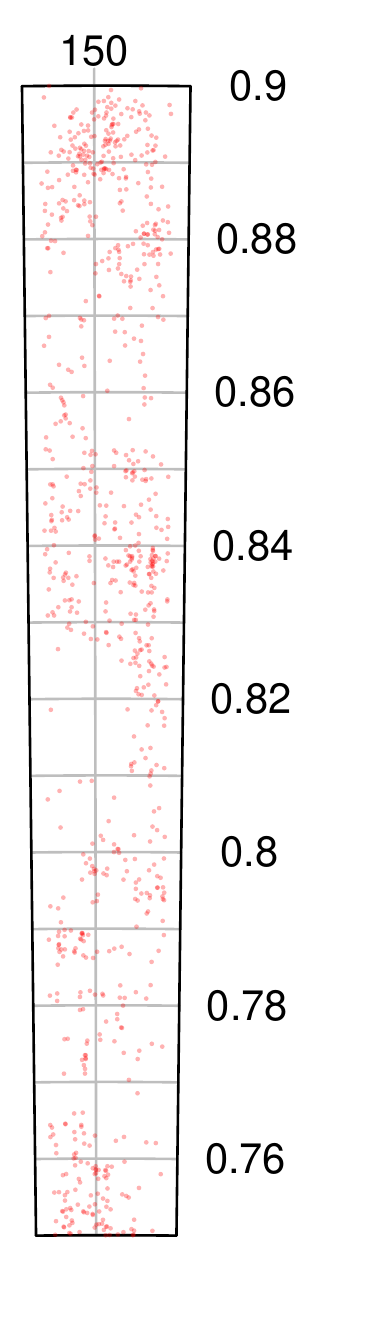}
\hspace{3mm}
\vline
\hspace{3mm}
\includegraphics[scale=0.6]{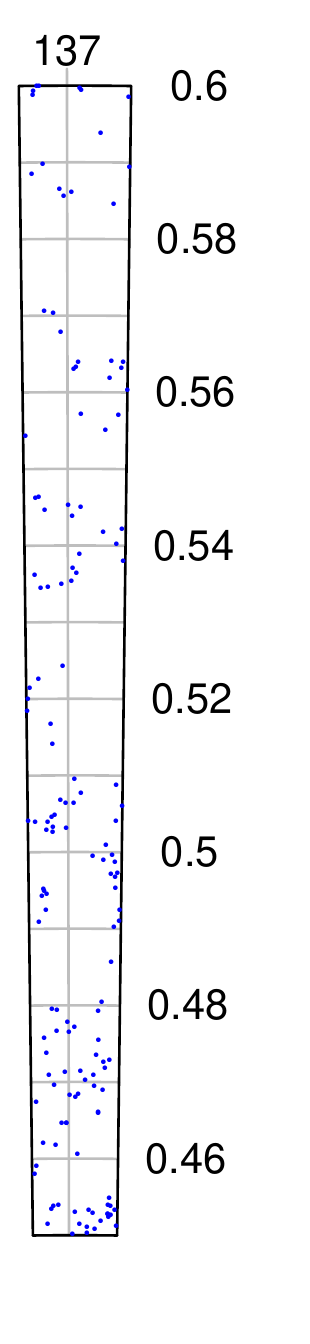}
\hspace{6mm}
\includegraphics[scale=0.6]{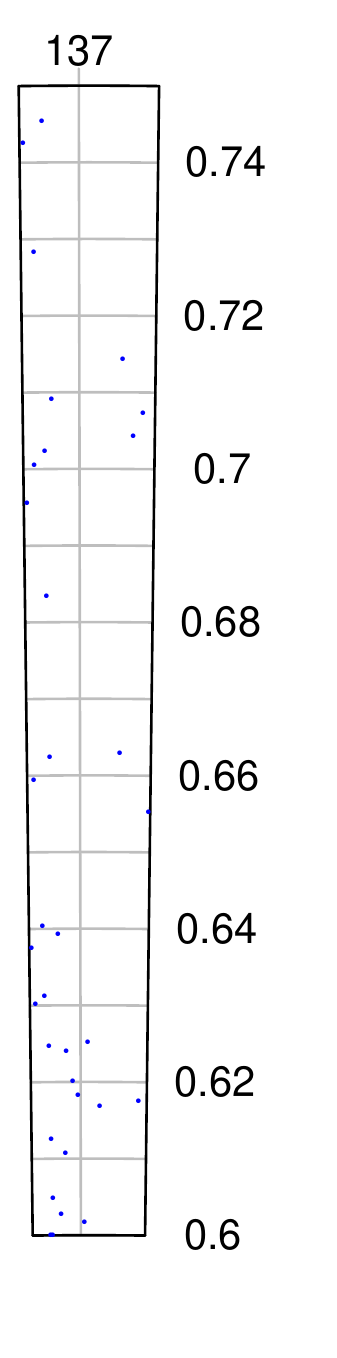}
\hspace{6mm}
\includegraphics[scale=0.6]{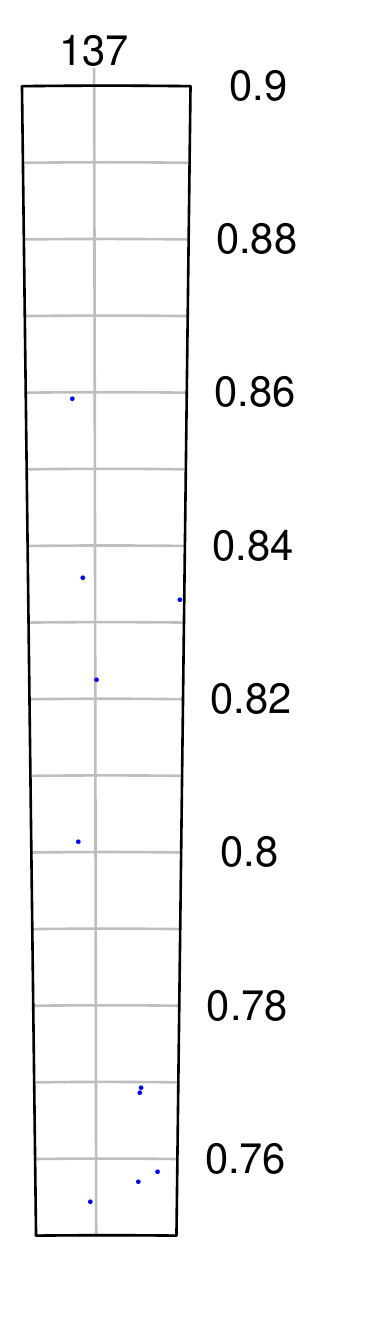}\\

\vspace{-6mm}

\includegraphics[scale=0.65]{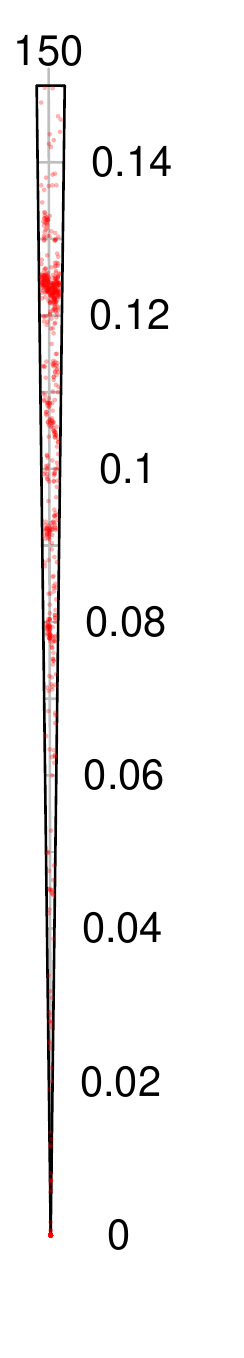}
\hspace{10mm}
\includegraphics[scale=0.65]{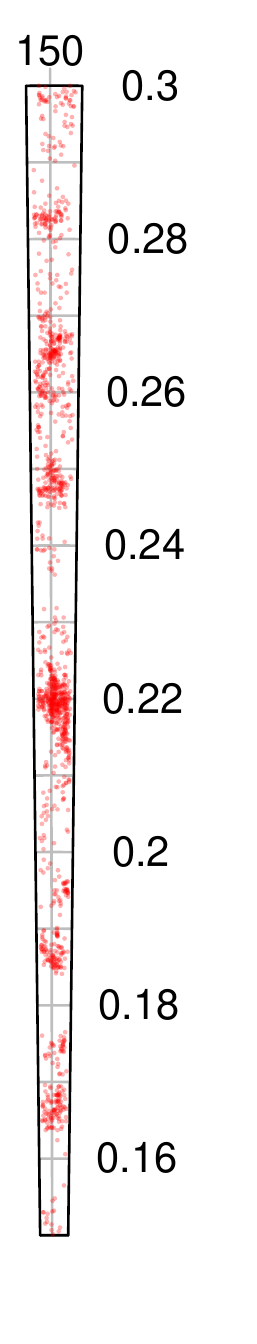}
\hspace{10mm}
\includegraphics[scale=0.65]{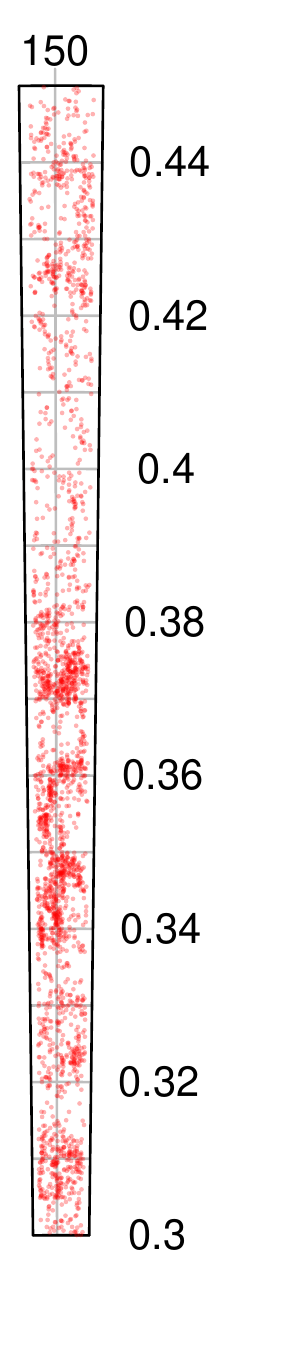}
\hspace{3mm}
\vline
\hspace{3mm}
\includegraphics[scale=0.65]{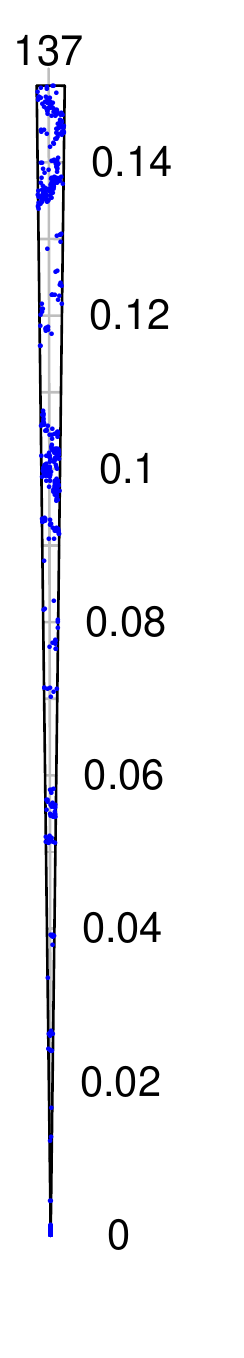}
\hspace{10mm}
\includegraphics[scale=0.65]{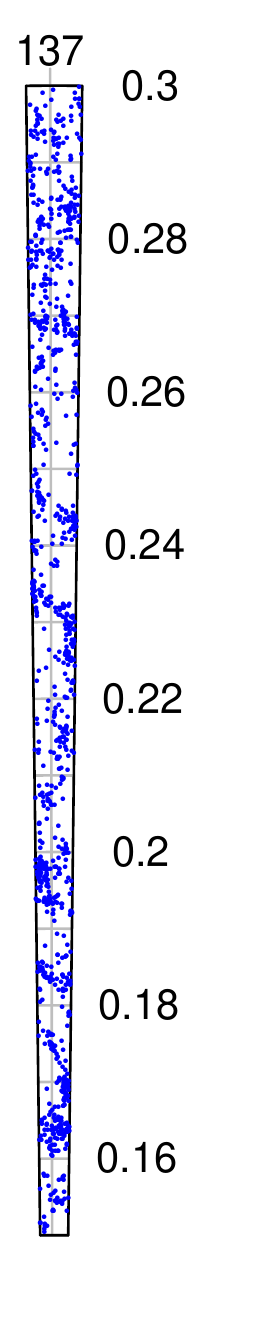}
\hspace{10mm}
\includegraphics[scale=0.65]{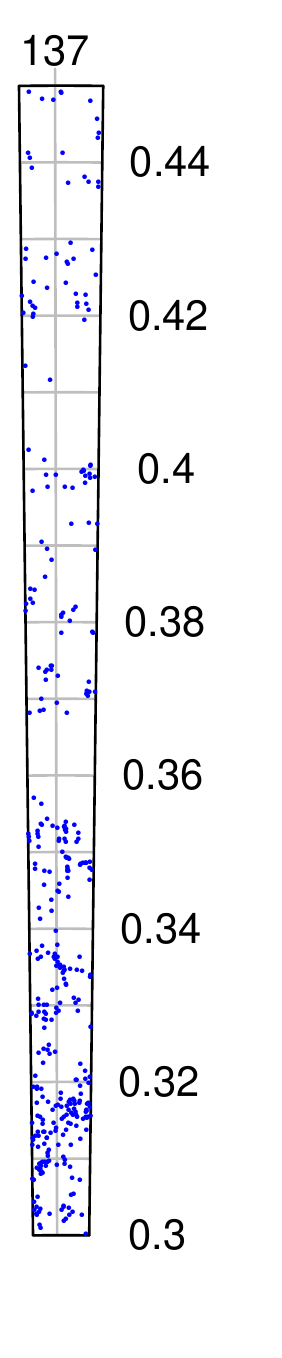}

\vspace{-4mm}

\caption{The light cone of sources in our G10-HR sample (left) in comparison to the same size volume in the GAMA-G09 region (right). Panels are split in to $\Delta z=0.15$ ranges, with redshift displayed on the vertical axis, in order to highlight structure in the galaxy distribution. }
\label{fig:light_cones}
\end{center}
\end{figure*}

\subsection{G10 and the GAMA low-$z$ sample}

\begin{figure}
\begin{center}

\includegraphics[scale=0.465]{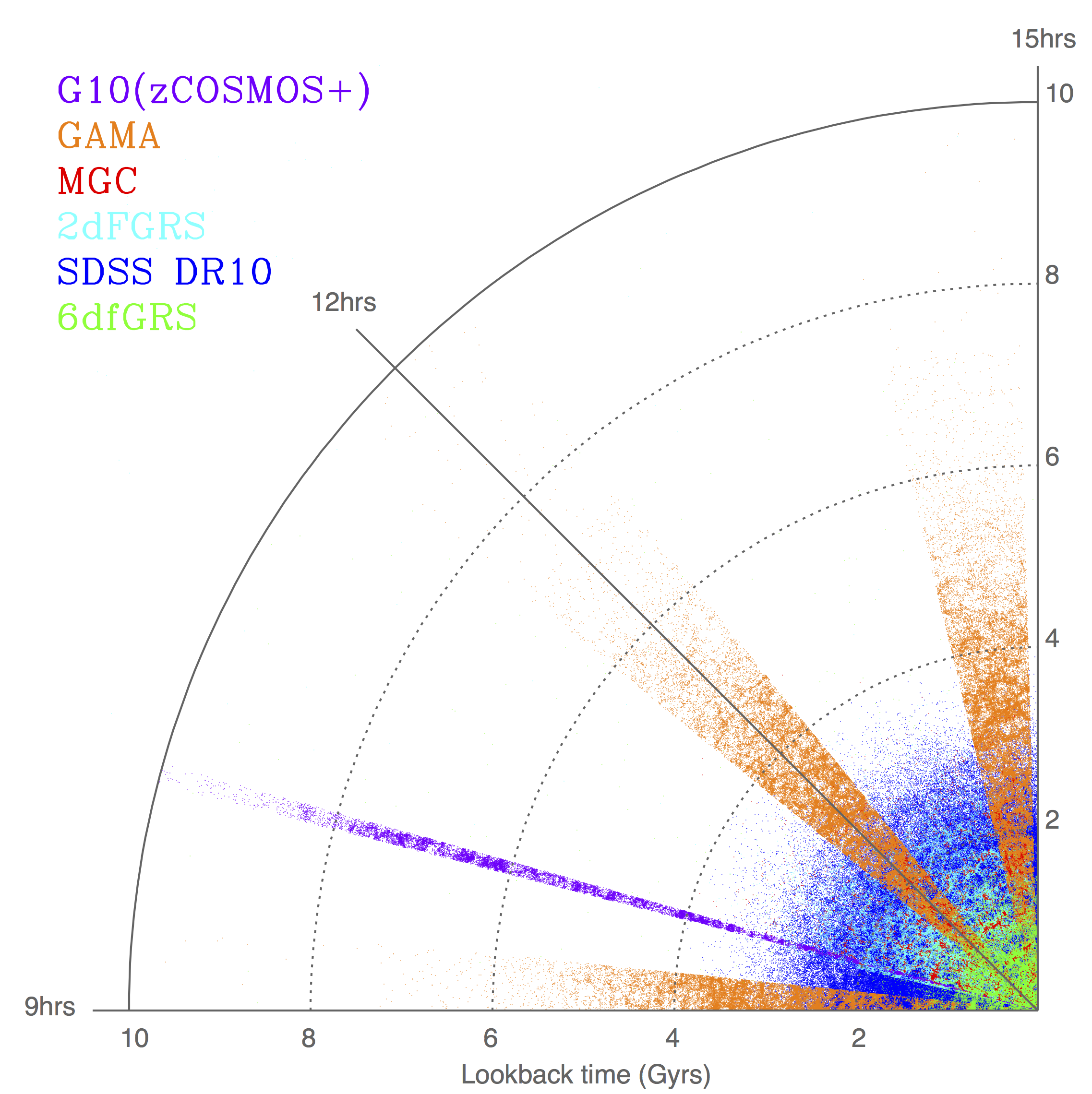}

\caption{The G10 region in comparison to other low redshift surveys. Only GAMA equatorial regions are shown (orange). The G10 region covers a much smaller area than the GAMA sample but has a higher source density and extends to higher redshift. }
\label{fig:gama_light_cones}
\end{center}
\end{figure}

In this section we compare our G10 catalogue with the GAMA II low redshift sample \citep{Liske14}. Figure \ref{fig:gama_light_cones} shows the light cones for the equatorial GAMA II regions (and other local redshift surveys) in comparison to the new G10 region. The G10 region covers a much smaller area than all other GAMA II regions ($\sim1$\,deg$^2$ in comparison to $\sim60$\,deg$^2$/per GAMA II region) but extends to much higher redshift (median $z\sim0.5$ in comparison to $z\sim0.2$ for GAMA II - see top panel Figure \ref{fig:Mr_dist}). In addition, the G10 region has a much higher source density of spectroscopically confirmed galaxies with $\sim$16,500 sources\,deg$^{-2}$ compared to $\sim$1000 sources\,deg$^{-2}$ in GAMA II. To draw direct comparisons Figure \ref{fig:light_cones} also shows light cones for a G10 sized volume from within the G09 region (right panels). The source density is much lower in the G09 volume and drops dramatically at $z>0.4$ - obviously this does not highlight the much larger volume, brighter magnitude limit and much higher completeness of the GAMA II survey in comparison to the G10 region.        

The deep VLT observations in the zCOSMOS-bright survey not only pushes the G10 sample to much higher redshifts but also fainter absolute magnitudes for a given redshift. Figure \ref{fig:Mr_dist} displays the absolute $r$- and $i$-band magnitude as a function of redshift for equatorial region GAMA II galaxies and our new spectroscopically confirmed sample in the COSMOS region (z\_use=1). We calculate absolute magnitudes using the cosmology outlined in Section \ref{sec:intro} and apply an approximate k-correction of

\begin{equation}
k(z)=0.053z+0.78z^2
\end{equation}

to account for the fact that these observations probe varying rest-frame spectral regions as a function of redshift.   

Considering the $r$-band, our new sample is on average $\sim$1.5 magnitudes fainter than GAMA at all redshifts and extends to $\gtrsim3$ magnitudes fainter at $z>0.4$.  The blue circles display the characteristic $i$-band magnitude, M$_{i}^{*}$, as a function of redshift taken from the Canada-France Hawaii Telescope Legacy Survey \citep[CFHTLS ,][]{Ramos11},  in the left panel this is scaled to M$_{r}^{*}$ using the median $r-i$ colour of a combined GAMA II + G10 sample at $\Delta z\pm0.1$ about the M$_{i}^{*}$ redshift. While the typical GAMA II galaxy is largely consistent with M$_{i}^{*}$ out to $z\sim0.4$, our new sample targets galaxies below M$_{i}^{*}$ out to $z<1.2$, probing the faint end of the galaxy luminosity function - and ultimately allowing the identification of much lower mass systems.

\subsection{The CHILES region}

As noted previously, the CHILES survey is currently obtaining deep HI observations at $0<z<0.45$ for a 0.3\,deg$^{2}$ FWHM region of the COSMOS region. The survey has been designed so that it will be able to detect, at the highest redshift, M$_{\mathrm{HI}}\sim3 \times10^{10}$\,M$_{\odot}$ at $5\sigma$, assuming a 150\,km\,s$^{-1}$ profile width. This sensitivity will allow for direct imaging in HI of at least 300 galaxies across the entire redshift range. The survey aims to reach an average rms of $\sim50\mu$Jy per 31\,kHz channel in a total observing time of 1000 hours.

\begin{figure}
\begin{center}
\includegraphics[scale=0.5]{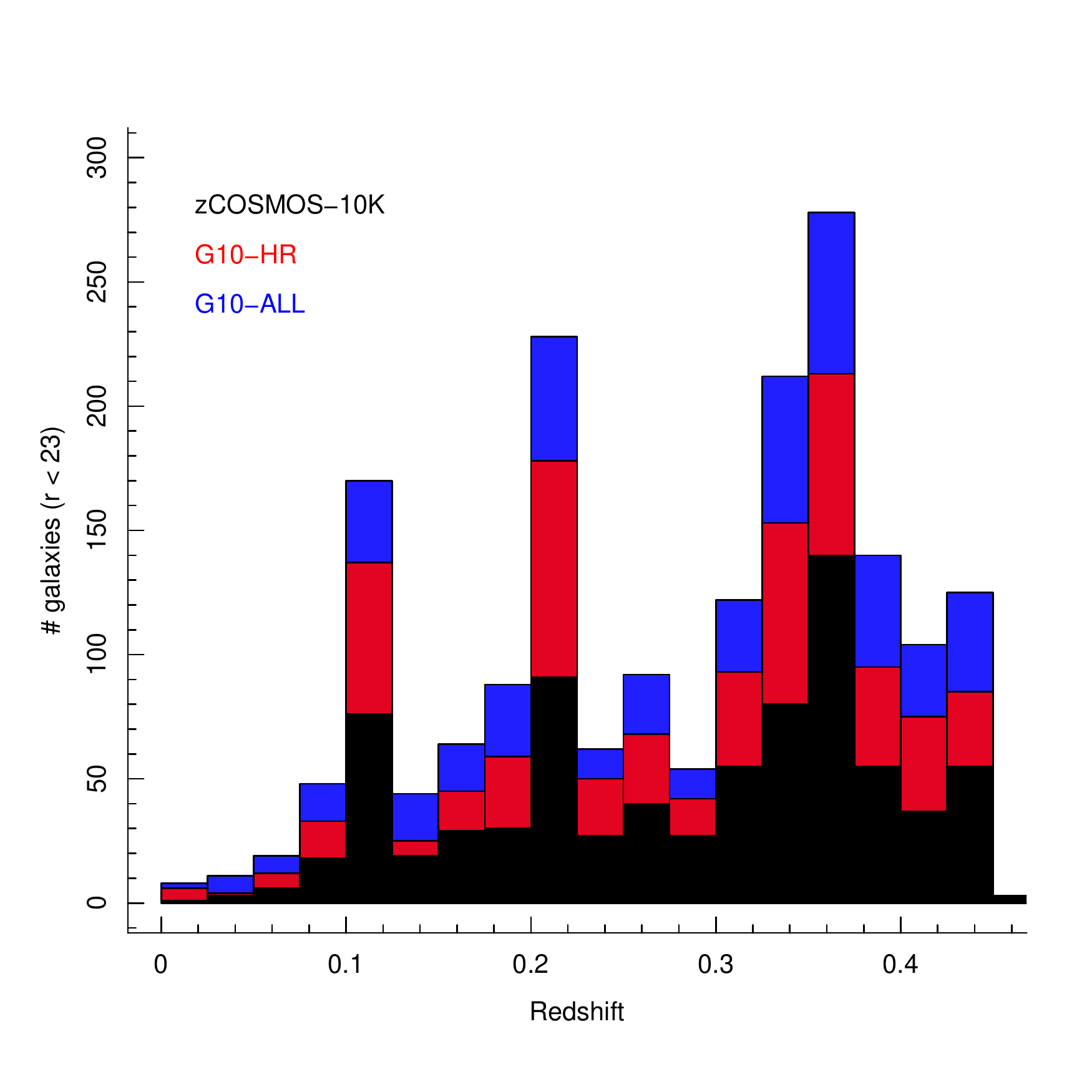}

\caption{The redshift distribution of galaxies in the CHILES volume. Our new G10 sample shows a significant increase in the number of spectroscopical confirmed galaxies in the CHILES region - required for HI stacking of non-detected sources.  }
\label{fig:CHILES_red}
\end{center}
\end{figure}

It is interesting to consider the number of robust G10 redshifts contained within the CHILES region - which may eventually be used for deep HI stacking, environmental measures and optically motivated source finding. Figure \ref{fig:CHILES_red} shows the redshift distribution of G10 sources in the CHILES volume in comparison to the zCOSMOS-10k sample. In total our G10-HR and G10-ALL samples contain 1373 and 1869 galaxies in the CHILES region respectively in comparison to just 597 in the publicly  available zCOSMOS-10k release.

\section{Conclusions and future use of the G10 sample}
\label{sec:conclusions}

We have constructed a spectroscopic sample of 17,446 robust galaxy redshifts in the COSMOS region by combining all publicly available spectroscopic observations in the region with our own re-reduction of the zCOSMOS-bright data. We assign the most robust and high quality redshift to all sources in the COSMOS region and match to all publicly available photometric data. We then define the central, high completeness region of COSMOS as the G10 region and compile a high resolution, $r<23$\,mag \&  $i<22$\,mag spectroscopic G10 sample of 9,861 sources (G10-HR) and high+low resolution spectroscopic G10 sample of 12,462 sources (G10-ALL). We discuss this sample in comparison to the publicly available zCOSMOS-bright sample, highlighting that we have gained a significant number of confirmed sources at all redshifts, while retaining the same target completeness. All of our sample, including the associated photometry, will be made publicly available through a web based tool (see Appendix for details).

The purpose of defining a magnitude limited spectroscopic catalogue in the COSMOS region is to perform future analysis in a consistent manner to the low redshift GAMA sample. The G10 catalogue will form the basis of many subsequent projects using the tools developed for GAMA low redshift analysis but extended to higher redshift. This will allow us to perform direct comparisons between the high redshift G10 sample and the extensive low redshift GAMA data, exploring the time evolution of individual galaxies and large scale structure. We will shortly produce our own independent and consistent matched aperture photometry for the full COSMOS region covering UV to FIR wavelengths using the Lambda Adaptive Multi-Band Detection Algorithm for R (LAMBDAR - Wright et al., in prep) and produce galaxy stellar masses using SED fits in a consistent manner to \cite[]{Taylor11}. Following this we will assign morphological classifications from the HST COSMOS data as in \cite{Kelvin14b} and define stellar mass functions as a function of redshift and morphological class. More extensive GAMA-type analysis will follow, such as bulge disc decompositions using Structural Investigation of Galaxies via Model Analysis \cite[SIGMA,][]{Kelvin12}, calculations of the evolution of  the galaxy mass-size relationship as a function of morphological type (see Lange et al. 2014), and studies of the Cosmic Spectral Energy distribution at intermediate redshifts \citep[$e.g.$][]{Driver12}.

In addition to these studies, we will ultimately also investigate the evolution of structure in the G10 region by producing group catalogues via numerous method, such as that discussed in \cite{Robotham11} and new techniques which are currently being developed (Kafle et al, in prep). Using the deep G10 spectroscopic catalogue we will be able to identify low mass halos at significantly higher redshift and investigate evolution in the halo mass function with time. However, it is as yet unclear as to whether the G10 sample is sufficiently complete to produce an accurate group catalogue (GAMA is $>98\%$ complete to $r<19.8$\,mag compared to just $\sim50\%$ to $r<23.0$\,mag in G10). Further spectroscopic observations may be required in the COSMOS region in order to increase the G10 completeness and provide a robust group catalogue with which to investigate the group distribution and other parameters constrained by the identification of linked galaxies - such as close-pair merger rates \cite[e.g.][]{Robotham14}.                     

\section*{Acknowledgements}

Firstly, we would like to thank L. Mannering for her construction of the web pages associated with the online \textsc{G10 Cutout Tool}. We would also like to acknowledge the help of the PRIMUS team and especially Alex Mendez for providing the P(z) distributions used in this work. Lastly, we would like to thank the anonymous referee for their helpful comments which led to significant improvements to this paper. GAMA is a joint European-Australasian project based around a spectroscopic campaign using the Anglo-Australian Telescope. The GAMA input catalogue is based on data taken from the Sloan Digital Sky Survey and the UKIRT Infrared Deep Sky Survey. Complementary imaging of the GAMA regions is being obtained by a number of independent survey programs including GALEX MIS, VST KiDS, VISTA VIKING, WISE, Herschel-ATLAS, GMRT and ASKAP providing UV to radio coverage. GAMA is funded by the STFC (UK), the ARC (Australia), the AAO, and the participating institutions. The GAMA website is {\btt http://www.gama-survey.org/}.

\appendix

\section{The G10 cutout tool and spectrum viewer}

The catalogue discussed in this paper is made publicly available through an online tool. The catalogue and associated products can be found here: \\
\\
{\btt http://ict.icrar.org/cutout/G10} \\

This catalogue is initially compiled from the 2007 photometric catalogue described in \cite{Capak07} and the updated 2008 COSMOS catalogue including the deep Subaru intermediate-band observations of \cite{Taniguchi07}, the 30-band photometric redshift catalogues of \cite{Ilbert09}, the zCOSMOS-10k bright catalogue \citep[][]{Lilly09}, PRIMUS catlaouge \citep[][]{Coil11,Cool13}, VVDS 10h region catalogue \citep[][]{Garilli08} and SDSS-DR10 \citep[][]{Ahn14}. We then also include additional parameters produced for the analysis in this paper.  The table contains all sources from \cite{Capak07} with the best available redshift associated with each source. All table parameters are fully documented in the associated $.par$ file and descriptions of the construction of the catalogue are provided in a $.notes$ file. Table \ref{tab:par} details the catalogue parameters not provided in the 2006 and 2008 COSMOS catalogues

\begin{table*}
\caption{G10/COSMOS catalogue parameters}
\begin{scriptsize}
\begin{center}
\begin{tabular}{l l l l l}
Name & Column & Units & UCD & Details \\ 
\hline 
\hline 

CATAID 	   &   		1 &	-	    &                 meta.id   	             &            G10 unique catalogue 6000000+ \\
*\_06 		   &	   2-46 & - & - & COSMOS 2006 catalogue parameters \\
*\_08 		   &	   47-127 & - & - & COSMOS 2008 catalogue parameters \\
ZP\_ILBERT      &    128 & -   &      src.redshift.phot         &    Best Photometric Redshift\\
STAR\_GALAXY\_CLASS   &129  & -    &     meta.code            &         Stellar flag  	  \\
ZL68\_ILBERT    &    130&  -      &   src.redshift.phot         &    lower limit, 68\% confidence level \\
ZU68\_ILBERT    &    131 & -     &    src.redshift.phot    &         upper limit, 68\% confidence level   \\
ZL99\_ILBERT    &    132  &-       &  src.redshift.phot        &     lower limit, 99\% confidence level    \\
ZU99\_ILBERT     &   133  &-     &    src.redshift.phot       &      upper limit, 99\% confidence level   \\
CHI2\_ILBERT    &    134  &-     &    stat.fit.residual       &      reduced chi2 for photometric redshift	\\		    
RA\_ZCOS        &    135  &deg    &   pos.eq.ra              &       RA (J2000.0) from zCOSMOS-bright 10k   \\
DEC\_ZCOS     &      136 & deg    &   pos.eq.dec          &          Dec (J2000.0) from zCOSMOS-bright 10k    \\
Z\_ZCOS        &     137  &-     &    src.redshift           &       Spectroscopic redshift from zCOSMOS-bright 10k  \\
ZQUALITY\_ZCOS   &   138 & -  &       meta.code      &              Quality flag zCOSMOS-bright 10k     \\
filename\_ZCOS &      139  &-       &  meta.id           &            Filename of zCOSMOS-bright 10k 1D reduced spectrum\\
RA\_PRIMUS     &     140  &deg    &   pos.eq.ra           &          RA (J2000.0) from PRIMUS     	\\		      
DEC\_PRIMUS    &     141 & deg    &   pos.eq.dec    &                Dec (J2000.0) from PRIMUS      \\
Z\_PRIMUS      &     142  &-      &   src.redshift       &           Spectroscopic redshift from PRIMUS  \\
ZQUALITY\_PRIMUS   & 143 & -      &   meta.code         &            Quality flag from PRIMUS \\
SLIT\_RA\_14      &   144  & deg   &    pos.eq.ra           &          RA (J2000.0) of slit position\\
SLIT\_DEC\_14       & 145  & deg    &   pos.eq.dec     &               Dec (J2000.0) of slit position \\
SLIT\_OFF          & 146 & arcsec  &  pos.angDistance        &       Offset between SLIT\_RA, SLIT\_DEC and RA\_06, DEC\_06\\
Z\_BEST      &       147 & -    &     src.redshift      &            The best-fit redshift \\
Z\_GEN          &    148  &-      &   meta.code          &           Numerical code quantifying redshift matches\\
Z\_USE          &    149  &-     &    meta.code       &              Numerical code indicating use of redshift\\
AUTOZ1         &    150  &-    &     src.redshift     &             Best-fit AUTOZ \\
AUTOZ1\_SIGMA     &  151 & -     &    stat.weight       &            Statistical significance of best-fit AUTOZ \\
AUTOZ1\_TEMP    &    152  &-     &    meta.id        &               Model galaxy template from best-fit AUTOZ \\
AUTOZ2       &      153 & -      &   src.redshift       &           2nd best-fit AUTOZ \\
AUTOZ2\_SIGMA     &  154 & -       &  stat.weight         &          Statistical significance of 2nd best-fit AUTOZ \\
AUTOZ2\_TEMP  &      155 & -      &   meta.id          &             Model galaxy template from 2nd best-fit AUTOZ \\
AUTOZ3       &      156  &-      &   src.redshift          &        3rd best-fit AUTOZ \\
AUTOZ3\_SIGMA    &   157 & -   &      stat.weight         &          Statistical significance of 3rd best-fit AUTOZ \\
AUTOZ3\_TEMP   &     158  &-        & meta.id           &            Model galaxy template from 3rd best-fit AUTOZ \\
AUTOZ4       &      159  &-       &  src.redshift            &      4th best-fit AUTOZ \\
AUTOZ4\_SIGMA     &  160&  -      &   stat.weight         &          Statistical significance of 4th best-fit AUTOZ \\
AUTOZ4\_TEMP     &   161 & -       &  meta.id         &              Model galaxy template from 4th best-fit AUTOZ \\
SPEC\_FILENAME   &   162  &-     &    meta.id        &               Filename of 1D reduced spectrum\\
Z\_EYE        &      163  &-        & src.redshift            &      Redshift from visual inspection 	\\
ZQUALITY\_EYE   &    164 & -     &    meta.code     &                Quality flag from visual inspection \\
R\_MAG\_BEST      &   165  &-     &    phot.mag;em.opt.R   &          Best R-band magnitude \\
QUALITY\_FLAG    &   166  &-     &    meta.code          &           Spectral quality flag\\
RA\_VVDS       &     167  &deg     &  pos.eq.ra           &          RA (J2000.0) from VVDS\\
DEC\_VVDS   &        168 & deg     &  pos.eq.dec             &       Dec (J2000.0) from VVDS \\
Z\_VVDS        &     169  &-       &  src.redshift      &            Spectroscopic redshift from VVDS\\
ZQUALITY\_VVDS   &   170 & -     &    meta.code     &                Quality flag from VVDS \\
RA\_SDSS      &      171  &deg    &   pos.eq.ra       &              RA (J2000.0) from SDSS-DR10    \\
DEC\_SDSS     &      172 & deg     &  pos.eq.dec       &              Dec (J2000.0) from SDSS-DR10  \\
CLASS\_SDSS   &      173 & deg    &   meta.code    &                 STAR/GALAXY class from SDSS-DR10   \\
Z\_SDSS          &   174 & -    &     src.redshift          &        Spectroscopic redshift from SDSS-DR10 \\
Z\_ERR\_SDSS    &     175 & -     &    src.redshift      &            Error on the spectroscopic redshift from SDSS-DR10 \\

\end{tabular}

\end{center}

\end{scriptsize}
\label{tab:par}
\end{table*}

\begin{figure}
\begin{center}
\includegraphics[scale=0.5]{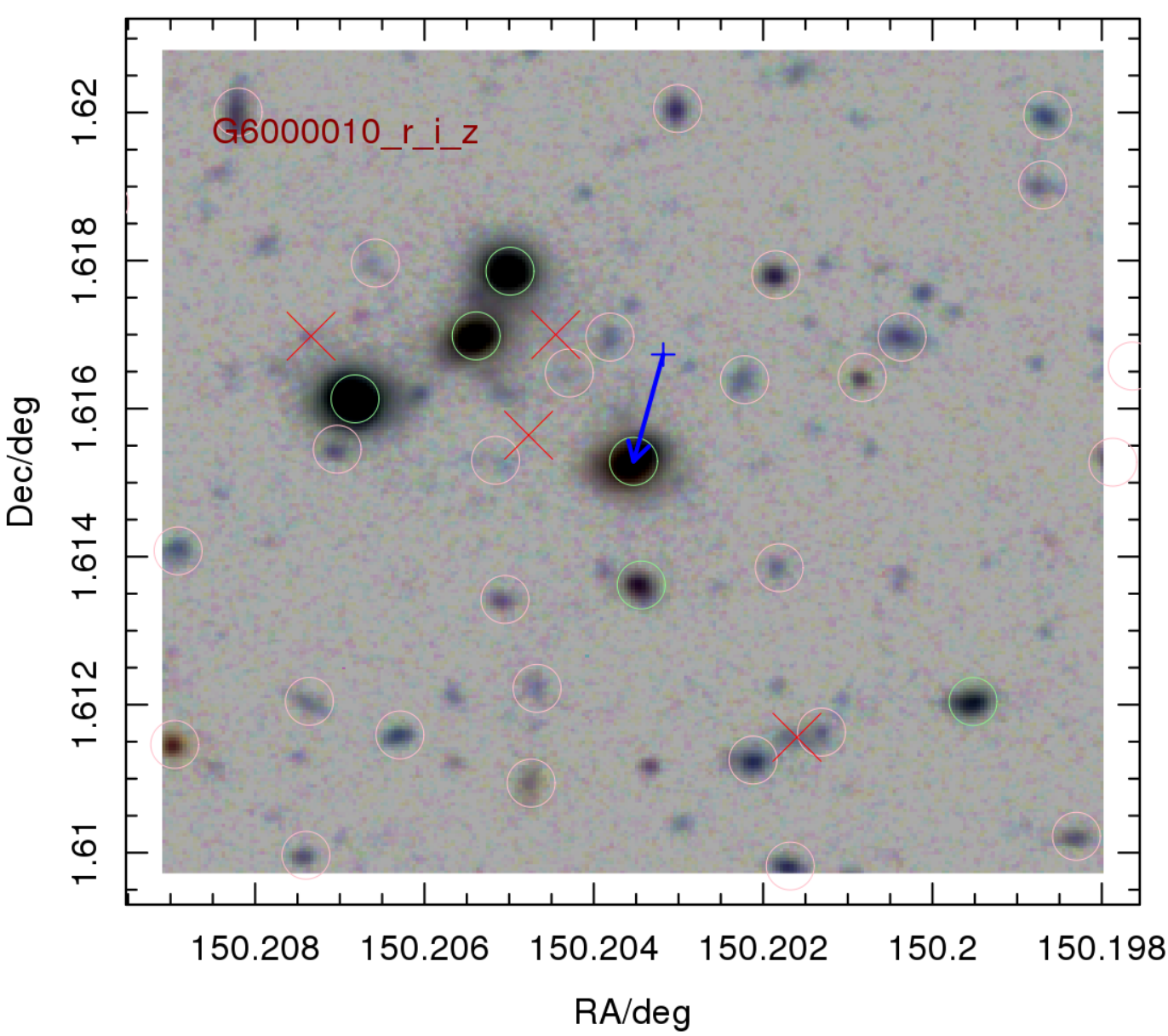}

\caption{Example of result from the G10 Cutout Tool displaying an automatically generated r, i, z RGB image for at $20\times20^{\prime\prime}$ region surrounding source CATADID=6000010. Overlaid are \citet{Capak07} sources. Green Circles = $i<23.5$ sources, Pink Circles = $i$>$23.5$ sources, red crosses = stars or sources which are fagged as bad, blue cross = zCOSMOS-bright slit position, and blue arrow=match between spectrum and photometric source. }
\label{fig:cutout}
\end{center}
\end{figure}

In addition to the raw catalogues, we also provide a tool for exploring the COSMOS data used in this work and the associated data produced as part of our current analysis. The \textsc{G10 Cutout Tool} allows users to select sources from the G10 catalogue via an ID tag, or RA and DEC position and obtain image cutouts for any combination of all available photometric data in both png and FITS format. These cutouts can also include overlays of \cite{Capak07} source positions and zCOSMOS-bright slit positions including the position matching discussed in this work. The tool allows for these cutout regions to be downloaded in FITS format for further analysis and will automatically generate RGB images and multi-band GIFs for any photometric data in the COSMOS region (for example see Figure \ref{fig:cutout}).

\begin{figure*}
\begin{center}
\includegraphics[scale=1.4]{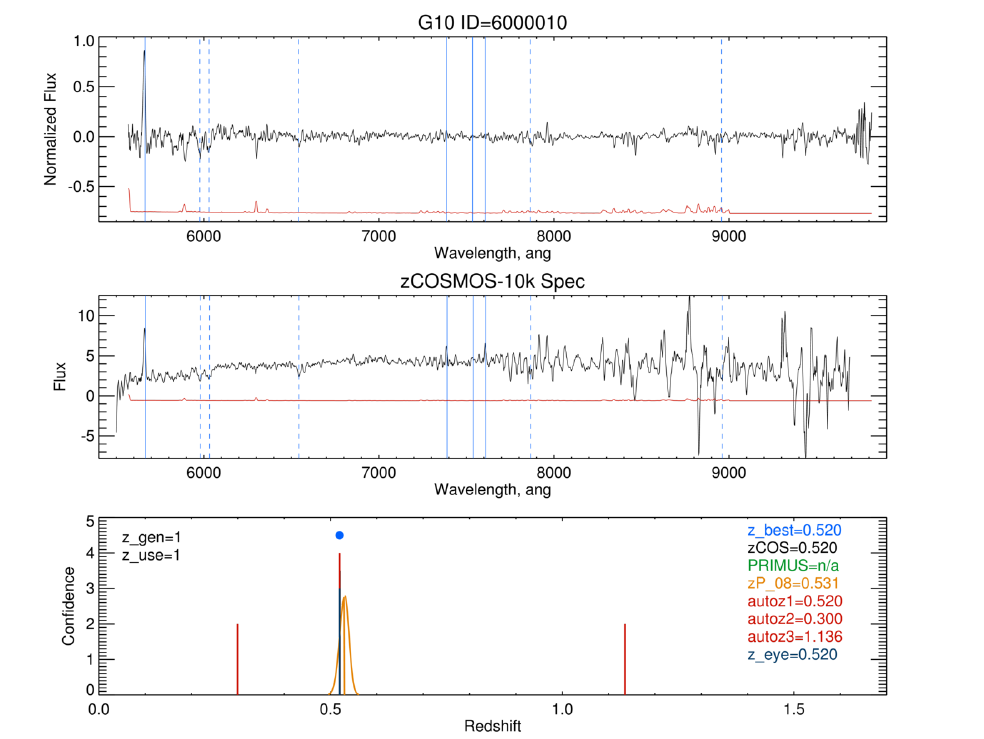}

\caption{Example of result from the G10 Cutout Tool displaying 1D spectra from our current re-reduction of the zCOSMOS-bright data, the zCOSMOS-10k release spectrum and a PDF of all available redshift information for source CATADID=6000010}
\label{fig:spec}
\end{center}
\end{figure*}

In addition to the photometric data, the tool provides plots of the source's continuum subtracted 1D spectrum from our re-reduction of the zCOSMOS data and the zCOSMOS-10k release spectrum (where available), with our best-fit redshift over plotted. Lastly, we provide a Probability Distribution Function (PDF) plot of all available redshifts for each source. These display redshifts from our current analysis, zCOSMOS-10k release, PRIMUS, zPhoto and our visual classifications. Figure \ref{fig:spec} displays and example of the spectrum/PDF plots provided with the photometric cutouts for source CATAID=6000010.  

Once further data/analysis of the G10 sample and COSMOS region becomes available, these will also be included in the online tool for public access. We ultimately aim to build a comprehensive online tool for the analysis of all sources in the COSMOS region.   

\begin{thebibliography}{}

\bibitem[Ahn et al.(2014)]{Ahn14} Ahn, C.~P., Alexandroff, 
R., Allende Prieto, C., et al.\ 2014, \apjs, 211, 17 


\bibitem[Alpaslan et al.(2012)]{Alpaslan12} Alpaslan, M., 
Robotham, A.~S.~G., Driver, S., et al.\ 2012, \mnras, 426, 2832 


\bibitem[Alpaslan et al.(2014)]{Alpaslan14} Alpaslan, M., 
Robotham, A.~S.~G., Driver, S., et al.\ 2014, \mnras, 438, 177

\bibitem[Baldry et al.(2010)]{Baldry10} Baldry, I.~K., Robotham, 
A.~S.~G., Hill, D.~T., et al.\ 2010, \mnras, 404, 86

\bibitem[Baldry et al.(2012)]{Baldry12} Baldry, I.~K., Driver, 
S.~P., Loveday, J., et al.\ 2012, \mnras, 421, 621 

\bibitem[Baldry et al.(2014)]{Baldry14} Baldry, I.~K., Alpaslan, 
M., Bauer, A.~E., et al.\ 2014, arXiv:1404.2626 

\bibitem[Bremer et al.(2010)]{Bremer10} Bremer, M.~N., Silk, J., 
Davies, L.~J.~M., \& Lehnert, M.~D.\ 2010, \mnras, 404, L69 

\bibitem[Brough et al.(2013)]{Brough13} Brough, S., Croom, S., 
Sharp, R., et al.\ 2013, \mnras, 435, 2903 

\bibitem[Capak et al.(2007)]{Capak07} Capak, P., Aussel, H., 
Ajiki, M., et al.\ 2007, \apjs, 172, 99 

\bibitem[Coil et al.(2011)]{Coil11} Coil, A.~L., Blanton, 
M.~R., Burles, S.~M., et al.\ 2011, \apj, 741, 8 

\bibitem[Cool et al.(2013)]{Cool13} Cool, R.~J., Moustakas, 
J., Blanton, M.~R., et al.\ 2013, \apj, 767, 118 

\bibitem[Driver et al.(2011)]{Driver11} Driver, S.~P., Hill, 
D.~T., Kelvin, L.~S., et al.\ 2011, \mnras, 413, 971 

\bibitem[Driver et al.(2012)]{Driver12} Driver, S.~P., Robotham, 
A.~S.~G., Kelvin, L., et al.\ 2012, \mnras, 427, 3244 


\bibitem[Fern{\'a}ndez et al.(2013)]{Fernandez13} Fern{\'a}ndez, 
X., van Gorkom, J.~H., Hess, K.~M., et al.\ 2013, arXiv:1303.2659 

\bibitem[Garilli et al.(2008)]{Garilli08} Garilli, B., Le F{\`e}vre, O., Guzzo, L., et al.\ 2008, \aap, 486, 683 

\bibitem[Hopkins et al.(2013)]{Hopkins13} Hopkins, A.~M., Driver, 
S.~P., Brough, S., et al.\ 2013, \mnras, 430, 2047 

\bibitem[Ilbert et al.(2009)]{Ilbert09} Ilbert, O., Capak, P., 
Salvato, M., et al.\ 2009, \apj, 690, 1236 


\bibitem[Kelvin et al.(2012)]{Kelvin12} Kelvin, L.~S., Driver, 
S.~P., Robotham, A.~S.~G., et al.\ 2012, \mnras, 421, 1007 

\bibitem[Kelvin et al.(2014a)]{Kelvin14} Kelvin, L.~S., Driver, 
S.~P., Robotham, A.~S.~G., et al.\ 2014a, arXiv:1401.1817 

\bibitem[Kelvin et al.(2014b)]{Kelvin14b} Kelvin, L.~S., Driver, 
S.~P., Robotham, A.~S.~G., et al.\ 2014b, arXiv:1407.7555 


\bibitem[Kova{\v c} et al.(2014)]{Kovac14} Kova{\v c}, K., 
Lilly, S.~J., Knobel, C., et al.\ 2014, \mnras, 438, 717 

\bibitem[Knobel et al.(2012)]{Knobel12} Knobel, C., Lilly, 
S.~J., Iovino, A., et al.\ 2012, \apj, 753, 121 

\bibitem[McCracken et al.(2012)]{McCracken12} McCracken, H.~J., Milvang-Jensen, B., Dunlop, J., et al.\ 2012, \aap, 544, A156 


\bibitem[Lange et al.(2014)]{Lange14} Lange, R.,  et al.\ 2014, MNRAS accepted

 


\bibitem[Lilly et al.(2007)]{Lilly07} Lilly, S.~J., Le 
F{\`e}vre, O., Renzini, A., et al.\ 2007, \apjs, 172, 70 

\bibitem[Lilly et al.(2009)]{Lilly09} Lilly, S.~J., Le Brun, 
V., Maier, C., et al.\ 2009, \apjs, 184, 218


\bibitem[Liske et al.(2014)]{Liske14} Liske, J., et al.\ 2014, \mnras, in prep


\bibitem[Ramos et al.(2011)]{Ramos11} Ramos, B.~H.~F., 
Pellegrini, P.~S., Benoist, C., et al.\ 2011, \aj, 142, 41 


\bibitem[Robotham et al.(2010)]{Robotham10} Robotham, A., Driver, 
S.~P., Norberg, P., et al.\ 2010, PASA, 27, 76 

\bibitem[Robotham et al.(2011)]{Robotham11} Robotham, A.~S.~G., 
Norberg, P., Driver, S.~P., et al.\ 2011, \mnras, 416, 2640

\bibitem[Robotham et al.(2012)]{Robotham12} Robotham, A.~S.~G., 
Baldry, I.~K., Bland-Hawthorn, J., et al.\ 2012, \mnras, 424, 1448 


\bibitem[Robotham et al.(2013)]{Robotham13} Robotham, A.~S.~G., 
Liske, J., Driver, S.~P., et al.\ 2013, \mnras, 431, 167


\bibitem[Robotham et al.(2014)]{Robotham14} Robotham, A.~S.~G., 
Driver, S.~P., Davies, L.~J.~M., et al.\ 2014, arXiv:1408.1476  

\bibitem[Scoville et al.(2007)]{Scoville07} Scoville, N., Aussel, 
H., Brusa, M., et al.\ 2007, \apjs, 172, 1 

\bibitem[Springob et al.(2005)]{Springob05} Springob, C.~M., 
Haynes, M.~P., \& Giovanelli, R.\ 2005, \apj, 621, 215 

\bibitem[Taniguchi et al.(2007)]{Taniguchi07} Taniguchi, Y., 
Scoville, N., Murayama, T., et al.\ 2007, \apjs, 172, 9 

\bibitem[Taylor et al.(2011)]{Taylor11} Taylor, E.~N., Hopkins, 
A.~M., Baldry, I.~K., et al.\ 2011, \mnras, 418, 1587 

\bibitem[Zwaan et al.(2005)]{Zwaan05} Zwaan, M.~A., Meyer, 
M.~J., Staveley-Smith, L., \& Webster, R.~L.\ 2005, \mnras, 359, L30 


\end{thebibliography}
\end{document}